\documentclass[fleqn,10pt]{olplainarticle}
\title{Virtual staining for 3D X-ray histology of bone implants}

\author[1]{Sarah C. Irvine}
\author[2,3]{Christian Lucas}
\author[2]{Diana Krüger}
\author[4]{Bianca Guedert}
\author[1]{Julian Moosmann}
\author[2,4]{Berit Zeller-Plumhoff}
\affil[1]{Institute of Materials Physics, Helmholtz-Zentrum Hereon, 21502 Geesthacht, Germany}
\affil[2]{Institute of Metallic Biomaterials, Helmholtz-Zentrum Hereon, 21502 Geesthacht, Germany}
\affil[3]{Bruker Daltonics SPR, 20251 Hamburg, Germany}
\affil[4]{Data-Driven Analysis and Design of Materials, University of Rostock, 18059 Rostock, Germany}


\begin{abstract}
Three-dimensional X-ray histology techniques offer a non-invasive alternative to conventional 2D histology, enabling volumetric imaging of biological tissues without the need for physical sectioning or chemical staining. However, the inherent greyscale image contrast of X-ray tomography limits its biochemical specificity compared to traditional histological stains. Within digital pathology, deep learning-based virtual staining has demonstrated utility in simulating stained appearances from label-free optical images. In this study, we extend virtual staining to the X-ray domain by applying cross-modality image translation to generate artificially stained slices from synchrotron-radiation-based micro-CT scans. Using over 50 co-registered image pairs of micro-CT and toluidine blue-stained histology from bone-implant samples, we trained a modified CycleGAN network tailored for limited paired data. Whole slide histology images were downsampled to match the voxel size of the CT data, with on-the-fly data augmentation for patch-based training. The model incorporates pixelwise supervision and greyscale consistency terms, producing histologically realistic colour outputs while preserving high-resolution structural detail. Our method outperformed Pix2Pix and standard CycleGAN baselines across SSIM, PSNR, and LPIPS metrics. Once trained, the model can be applied to full CT volumes to generate virtually stained 3D datasets, enhancing interpretability without additional sample preparation. While features such as new bone formation were able to be reproduced, some variability in the depiction of implant degradation layers highlights the need for further training data and refinement. This work introduces virtual staining to 3D X-ray imaging and offers a scalable route for chemically informative, label-free tissue characterisation in biomedical research.

\end{abstract}

\begin{document}

\flushbottom
\maketitle
\thispagestyle{empty}

\section{Introduction}
\label{section:introduction}

Histological analysis is widely regarded within clinical pathology and biomedical research as the benchmark technique for characterising tissue architecture and cellular detail. Through chemical staining of thin sections and brightfield microscopy, conventional histology provides high specificity in identifying cell types, pathological states, and tissue organisation. However, this process is inherently destructive, labour-intensive, and fundamentally two-dimensional \citep{abraham_current_2024}. As a result, conventional histology captures only limited spatial context, posing significant challenges for studying complex 3D tissue and biomaterial interactions, particularly in large or heterogeneous specimens such as whole organs or biomedical implants.

Overcoming those limitations, three-dimensional (3D) X-ray imaging techniques such as synchrotron-radiation X-ray micro-tomography (SR-µCT) and high-resolution laboratory µCT present as powerful alternatives. These methods enable label-free, non-invasive visualisation of soft and mineralised tissues at micrometre-scale resolution. When applied correlatively with histology for reference, such approaches may be collectively termed `3D X-ray histology' or `X-ray virtual histology', providing volumetric datasets that retain cellular- and tissue-level detail across intact specimens \citep{albers_x-ray-based_2018, topperwien_three-dimensional_2018, rodgers_3d_2023}. However, despite their advantages, a central limitation of 3D X-ray histology is the lack of biochemical specificity: image contrast is intrinsically greyscale, derived from differences in X-ray attenuation or phase shift, and does not replicate the cell-type or structure-specific colour-coded information provided by traditional histological stains. While X-ray contrast agents based on high-\textit{Z} elements can be applied to enhance soft tissue visibility, these typically bind to tissue components differently than conventional stains, and the issue of bio-specificity remains unresolved. Some progress has been made through the development of modified histological stains that incorporate heavy metal complexes, offering improved compatibility with traditional staining protocols \citep{muller_nucleus-specific_2018, petzold_development_2024}. However, the heavy-metal related issues of toxicity, environmental safety, and clinical applicability are of some concern. As a result, computational methods that enhance the interpretability of raw X-ray volumes without the need for physical staining represent a valuable and increasingly important advancement. In one early X-ray histology study \citep{khimchenko_extending_2016}, a joint histogram-based analysis was used to map greyscale CT intensities to histological colour values, enabling stain-like colourisation of 3D volumes. However, this approach has seen limited uptake in subsequent research, potentially due to challenges in distinguishing overlapping attenuation profiles and the absence of spatial or contextual cues needed for more robust tissue differentiation.

Recent advances in deep learning, in particular the development of generative adversarial networks (GANs), have transformed the field of computational image synthesis, including applications of digital pathology. GANs are composed of two competing neural networks: a generator that synthesises images and a discriminator that evaluates their realism, trained adversarially to produce outputs indistinguishable from ground-truth data \citep{goodfellow_generative_2014}. In medical imaging, GANs have been applied to tasks such as segmentation, super-resolution, denoising, artefact reduction, and notably, cross-modality image-to-image translation, where a transformation is learned between different imaging modalities \citep{yi_generative_2019}. Examples include synthesising computed tomography (CT) from magnetic resonance imaging (MRI) or positron emission tomography (PET) from CT, which are clinically valuable when certain modalities are limited by cost, radiation exposure, or accessibility.

`Virtual staining' is a subset of GAN-based applications demonstrating a rapid uptake in digital pathology whereby label-free microscopy images are transformed into the visual equivalent of chemically stained histological images, for enhanced interpretability without the need for physical staining \citep{latonen_virtual_2024}. These techniques have demonstrated high-fidelity image-to-image translation in 2D, applied within various optical imaging modalities including autofluorescence \citep{rivenson_virtual_2019}, photo-acoustic microscopy \citep{kang_deep_2022} and quantitative phase imaging (QPI) \citep{abraham_mode-mapping_2022}, as well as standard brightfield microscopy \citep{koivukoski_unstained_2023}. However, to date virtual staining has not yet been extended to the X-ray domain, where the image formation mechanisms and contrast properties differ significantly. In this study, we introduce a deep-learning model for virtual staining in 3D X-ray histology, enabling direct synthesis of virtually stained 3D tissue volumes from raw greyscale µCT scans. This combination provides both the structural integrity of X-ray imaging and the interpretability of conventional histology.

For the base model architecture, we consider as reference two foundational models in image-to-image translation: Pix2Pix \citep{isola_image--image_2018} and CycleGAN \citep{zhu_unpaired_2017}, representing supervised and unsupervised approaches, respectively. Supervised models like Pix2Pix require spatially aligned paired datasets, which are often difficult or even impossible to obtain in medical imaging due to motion, resolution mismatch, or other acquisition constraints. CycleGAN addresses this limitation through a cycle-consistency loss that enables training with unpaired data. In earlier applications of GANs in medical imaging, Pix2Pix, CycleGAN and variants thereof comprised a majority of cross-modality synthesis tasks \citep{yi_generative_2019}, although as the field has progressed, there has been a shift towards increasingly task-specific GAN models tailored to address the unique challenges of modality translation and clinical relevance \citep{heng_survey_2024}. CycleGAN remains the most widely adopted unsupervised approach in virtual staining, where fully aligned datasets are rarely available \citep{latonen_virtual_2024}. In the related context of virtual stain transfer, where one histochemical stain is virtually transformed into another, CycleGAN was found to outperform other architectures, including Pix2Pix, even in settings where simulated paired data were available \citep{zingman_comparative_2023}.

Pix2Pix may still yield superior results when perfect alignment between modalities is available, whereas the unsupervised CycleGAN is ideal for situations with no assumed alignment. Many real-world medical imaging scenarios, however, fall between these extremes, with some degree of partial misalignment. This intermediate setting has motivated adaptations of CycleGAN for use with paired data, retaining its robust architecture while benefiting from supervision. The added stability offered by cycle-consistency constraints encourages learning of mappings that are reversible and coherent across imperfectly aligned image domains \citep{kaji_overview_2019}. Several studies have explored such paired CycleGAN frameworks, including work on cone-beam CT correction and dual-energy chest X-ray imaging, where misaligned data are actively corrected \citep{harms_paired_2019, ueda_deep_2025}, as well as MRI-to-CT synthesis \citep{lei_mrionly_2019} and contrast-enhanced mammography \citep{rofena_deep_2024}. These studies illustrate the flexibility of CycleGAN variants in modality translation tasks where precise pixel-wise alignment is lacking but a high level of spatial correspondence is preserved. Our work adopts a similar approach, employing a paired CycleGAN framework adapted to the X-ray and histology domains, and incorporating additional supervisory loss terms to improve structural fidelity in domain mapping. We evaluate the performance of this modified CycleGAN in comparison to both the standard CycleGAN and Pix2Pix models.

\begin{figure}[ht]
\centering
\includegraphics[width=0.9\linewidth]{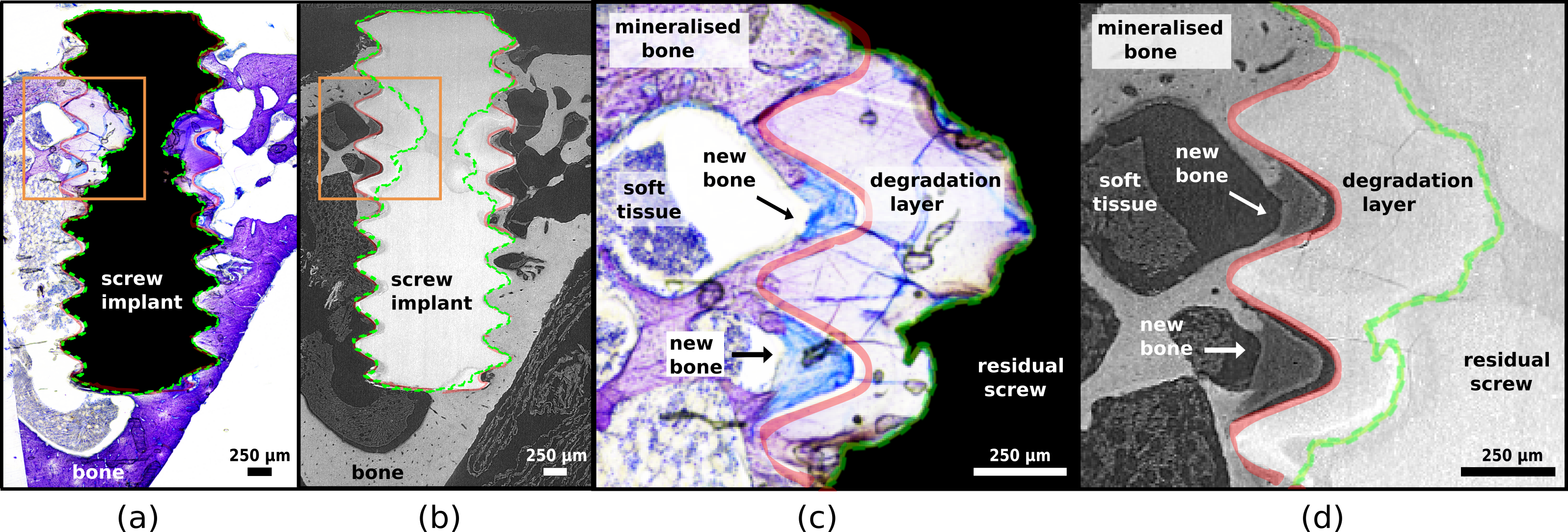}
\caption{ An illustrative example of our X-ray-histology application, with (real) paired toluidine-blue stained histological (a) and X-ray tomographic (b) slice images of a Mg-10Gd screw implant in bone extracted after 8 weeks of healing time in vivo. In the histology, the green line marks the interface between the residual alloy and the degradation layer. In the tomographic image, the red line indicates the boundary between the degradation layer and surrounding bone. The residual alloy appears black in histology because the metal blocks all light transmission. Regions identified as the degradation layer in histology correspond to areas with differing X-ray attenuation in SR-µCT to the alloy material. Higher magnification views (c, d) indicated by the orange boxes in both modalities, highlight the areas surrounding the degradation layer. Arrows in black (c) or white (d) point to regions of new (woven) bone, which in the histology appear blue in colour, and in µCT exhibit lower attenuation values due to lower mineralisation content. Figure adapted from \citet{kruger_high-resolution_2022} under a CC-BY-4.0 license.}
\label{fig:comparisonCThisto}
\end{figure}

We apply our methodology to the context of biodegradable magnesium-based implants in bone tissue, a challenging biomedical application that demands high-resolution, multi-scale techniques to assess material degradation and osseointegration \citep{zeller-plumhoff_utilizing_2021}. In previous multi-modal studies \citep{kruger_assessing_2021, sefa_multiscale_2023, iskhakova_multi-modal_2024}, we demonstrated the utility of correlative imaging to characterise complex tissue–material interactions, using a range of electron and X-ray microscopy techniques alongside conventional histology. In particular within the X-ray histology application of \citet{kruger_high-resolution_2022}, comparisons of bone-implant samples at different healing intervals revealed that magnesium alloys gradually form a stable degradation layer, surrounded by newly-formed non-mineralised bone. While histology confirmed the presence of these features, SR-µCT combined with a U-Net convolutional neural network (CNN) \citep{baltruschat_scaling_2021} enabled their tomographic segmentation, allowing 3D visualisation and volume-based quantification of key parameters such as bone–implant contact and degradation rate. Although corresponding 2D histology measurements and 3D µCT measurements were correlated, a difference in the details highlights the application as a strong potential candidate to benefit from 3D virtual staining.
Accordingly, to facilitate supervised learning, we collated and co-registered paired datasets from such prior studies, comprising upwards of 50 SR-µCT slices and their corresponding stained histological sections. An exemplary pair is shown in Figure \ref{fig:comparisonCThisto}, which is annotated to show key sample features that will be referenced in this text. The results of this application represent the first known demonstration of deep-learning-based virtual staining in 3D X-ray histology, bridging structural and chemical imaging modalities and enabling scalable, non-destructive analysis of tissue-implant interactions in 3D. 

\section{Methods and Materials}
\label{section:Methods-Materials}

Collected imaging datasets from the aforementioned correlative characterisation studies of biodegradable metal bone implants were re-used in this computational project, pertaining to samples with explants of magnesium-gadolinium screws (Mg-5wt.\%Gd and Mg-10wt.\%Gd), titanium (Ti) or polyether-ether-kethone (PEEK) screws implanted into Sprague Dawley rat tibia with healing periods of 4, 8 or 12 weeks. The Ti and PEEK screws were used as control materials to evaluate osseointegration without degradation of the implant. Animal experiments were conducted after ethical approval by the ethical committee at the Malmo/Lund regional board for animal research, Swedish Board of Agriculture (approval number DNR M 188-15). For a comprehensive description of the sample methodology, including the initial alloy production and animal study details, please refer to \citet{kruger_high-resolution_2022} and references therein.

\begin{figure}[ht]
\centering
\includegraphics[width=0.8\linewidth]{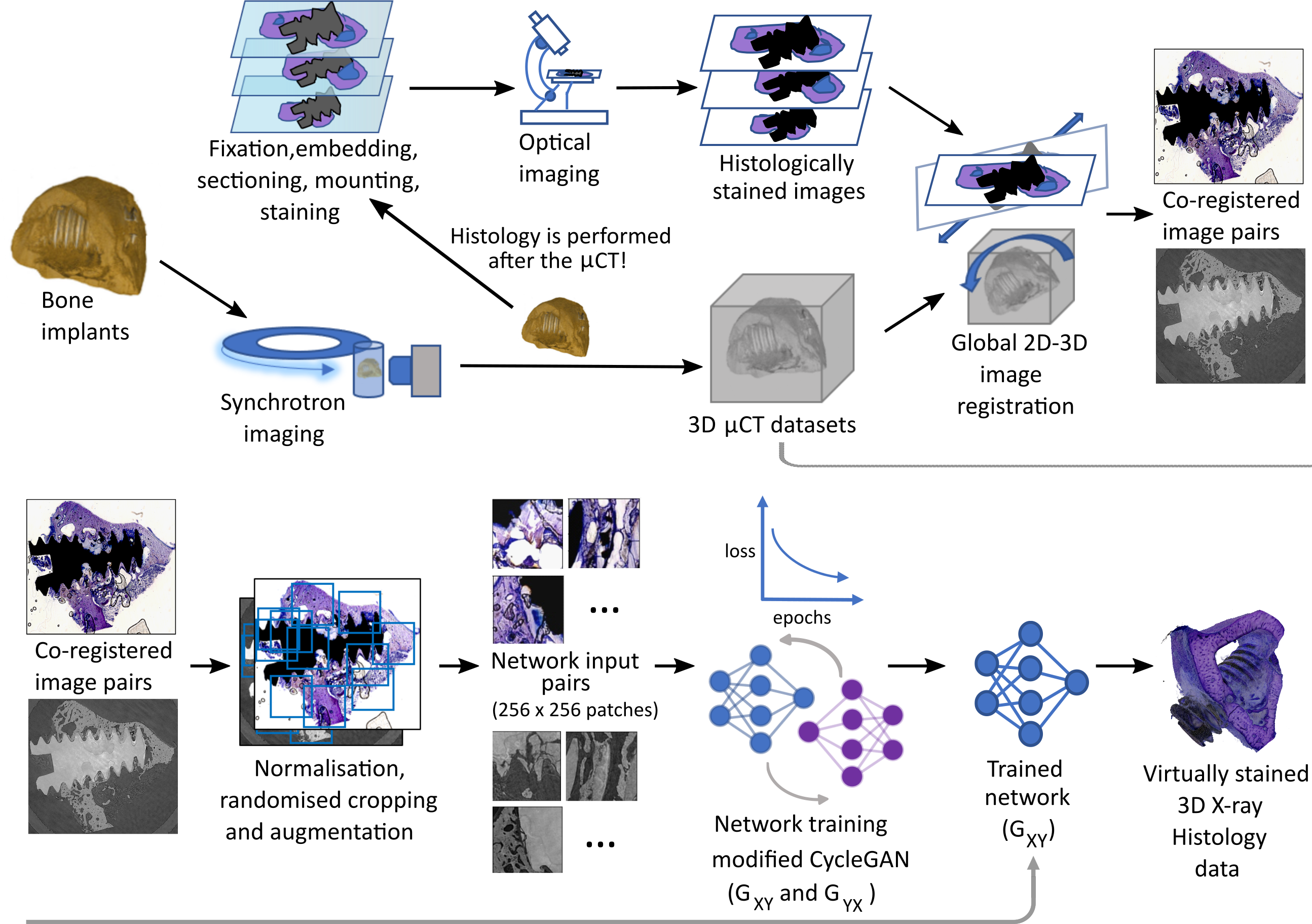}
\caption{Schematic of the methodology for 3D virtual staining in X-ray histology of bone implants, using a paired CycleGAN network.}
\label{fig:schematic-methodology}
\end{figure}

Following the sample preparation of the bone-implant block, the methodology timeline is illustrated schematically in Figure \ref{fig:schematic-methodology} which comprises the two imaging modalities followed by global registration to create the image pairs required for the supervised GAN inputs. Note that the histology must be performed after the tomography and not before, due to the destructive nature of the histological sectioning process. The second half of the methodology describes the investigated models, the data and its treatment including normalisation, augmentation, and use of a sample correspondence mask, as well as the metrics applied for comparative 2D analysis. Finally, we demonstrate the trained model in a 3D application by processing a stack of µCT slices into a virtually stained 3D X-ray histology volume.

\subsection{Sample preparation}
\label{subsection:SamplePreparation}

For the SR-µCT imaging, samples consisted of bone-implant blocks comprising screw (4 mm length, 2 mm in diameter and M2 thread) and surrounding bone tissue, explanted with a trephine bur of 6 mm diameter. The bone-implant blocks were fixed in 70$\%$ ethanol for 1 or more days and then dehydrated in a graded series of ethanol (samples were critically point dried).
The dried blocks were put into a standard Eppendorf tube (with arbitrary orientation) before being placed in the SR X-ray beam for tomographic imaging.

For the histology, which occurred after the SR-µCT imaging, explants were re-infiltrated with absolute ethanol and then embedded in methyl methacrylate resin by LLS Rowiak LaserLabSolutions GmbH (Hanover, Germany). Each sample was cut in half along the screw longitudinal axis with an Exakt saw, and then prepared for non-decalcified histology with the cutting-grinding technique \textit{ad modum} \citet{donath_preparation_1988}. Sections of about 40 µm were obtained, mounted on glass slides and stained (Histlab, Göteborg, Sweden) with a solution of Toluidine Blue-Pyronine Y. This is a combination aqueous solution of the two cationic dyes in which Toluidine Blue binds to acidic structures (e.g., nucleic acids) and Pyronin Y preferentially stains RNA (in red). Toluidine blue solution is widely used as a staining method for undecalcified bone tissue, allowing for identification of the mineralised bone matrix, osteoids, and soft tissues \citep{peev_modified_2024}. When combined with pyronin Y it results in staining of the bone tissue in various shades of purple (darker corresponding to younger bone) while the soft tissue is stained with more of a blue tint \citep{kropatsch_quantification_2007}. New woven bone at the degradation layer interface may also be stained bright blue \citep{kruger_high-resolution_2022}. 

As a smaller secondary dataset, we also collated histology samples that were stained with Hematoxylin and Eosin (H\&E) (LLS Rowiak, Hannover Germany). Sections were laser-cut by LLS Rowiak in a process which excludes the residual screw implant from the final mounted tissue sections (whilst retaining a portion of the screw degradation layer material). H \&E is the most widely used stain across all cell types; in bone it results in a deep pink, almost red stain for the mineralised tissue and purple staining of the soft tissue. Examples of this less complete dataset are presented in the Supplementary Information Section \ref{sec:si_HE}. 

\subsection{Imaging}
\label{subsection:Imaging}

The synchrotron-radiation micro-tomography datasets were acquired over several beamtimes at the imaging beamline (IBL) P05, which is operated by the Helmholtz-Zentrum Hereon, at the PETRA III storage ring of the Deutsches Elektronen Synchrotron (DESY) in Hamburg, Germany \citep{wilde_micro-ct_2016}. The primary form of contrast is attenuation-based, with minimal in-line phase contrast, acquired in full-field transmission geometry. As the beamtimes were part of different studies, a range of various acquisition parameters were utilised, including X-ray energies (ranging from 25 to 45 keV to accommodate the attenuation range of the different screw alloy materials, i.e., PEEK, Mg and Ti) and cameras (CCD vs CMOS) coupled to the X-ray microscope. Further details are given within \citet{kruger_high-resolution_2022}. Volumes of approximately 6 mm $\times$ 6 mm $\times$ 6 mm were scanned, with effective pixel sizes of either 1.2 or 2.4 µm. Tomograms were reconstructed using a MATLAB based framework and via the ASTRA toolbox for tomographic backprojection. Following tomographic reconstruction all datasets were downsampled to a voxel size of 5 µm (1.3k $\times$ 1.3k pixel slices). 

The collected histological images were also obtained over multiple measurement periods. All stained sections were imaged with one of two white light optical microscopes coupled to a camera (Optical microscope 1: the Nikon Eclipse Ci-L and DS-Fi3 camera controlled by NIS-Elements software, Tokyo, Japan. Optical microscope 2: the Zeiss AxioCam MrC controlled by AxioVision software, Oberkochen, Germany). Images were acquired with nominal magnification of 10$\times$ in wide field mode with an effective pixel size of 0.44 µm, and whole slide imaging (WSI) was achieved via guided manual stepping of the sample stage together with the inbuilt automatic image tiling function. Some sample images may be described as a partial WSI centered on the screw (approximate field of view is 2.5 $\times$ 4.5 mm), whilst others contain the full extent of sectioned screw plus bone area (roughly 6 mm $\times$ 6 mm). There is no consistent final array size with typical dimensions ranging from 6k to 13k pixels. Following registration (described in the next section), the transformed WSI images were downscaled to match the pixel size of the µCT slice pair (5 µm).

\subsection{Registration}
\label{subsection:Registration}
SR-µCT and histology WSI image pairs were co-registered using a semi-automated global 2D-3D registration pipeline outlined in \citet{irvine_multi-modal_2024}. In an iterative process based on minimising the mutual information metric, a 3D rigid transformation is applied to the µCT dataset to find the optimal virtual plane fit to the histology image. This is followed by an affine transformation of the histology image, incorporating non-uniform scaling and local shear to reverse such (minimal) distortion which may have occurred during histological sample preparation after the tomography measurement. The pixel size of the transformed output registered histology/µCT image pair is the same as the input µCT dataset.
Finally, both transformed slices were cropped to fit the minimum field of view of the pair.

\subsection{Datasets}
\label{subsubsection:datasets}

A total of 53 co-registered µCT and Toluidine-Blue stained histology WSI pairs were collated for this project. These comprised 38 samples containing Mg (-based implants), 5 containing Ti and 10 containing PEEK. We performed a 5-fold cross-validation with a 40 /10 train/validation split rotated through 5 times. Each validation set was randomly selected whilst keeping a representative 7:1:2 ratio of Mg:Ti:PEEK samples. The 3 WSI pairs for testing (comprising 2 Mg and 1 PEEK) were kept separate and never involved in training in order to prevent data leakage.

Corresponding details of the secondary H\&E-stained dataset are given in the Supplementary Subsection \ref{subsection:si_HEdatasets}.

\subsection{Normalisation}
\label{subsubsection:normalisation}

Normalisation was applied to all histology and µCT image pairs to mitigate domain shift. As the data were compiled from multiple measurement sessions and datasets, they exhibit substantial variation in image quality and statistics. For both modalities, regions of interest (ROIs) of both rigid bone and air/background were sampled to define scaling bounds. With the µCT data, intensities were linearly scaled to set the mean values of bone and background (30,000 and 10,000 respectively, for a 16-bit greyscale image file). For the histology RGB images, a simple white-balance correction was applied via the mean sampled background value $\mu_{R,G,B}$ (such that $\mu_R=\mu_G=\mu_B$), followed by contrast-stretching. Distribution bounds were defined by the 1st percentile of the bone ROI intensity distribution $i$ ($f_{\min} = \min_{R,G,B}(i) - c$, clamped to 10) and the 99th percentile of the background ROI distribution $j$ ($f_{\max} = \max_{R,G,B}(j) + c'$, clamped to 255 for 8-bit images). The same constants $c$ and $c'$ were applied across all channels to preserve chromaticity.

\subsection{Network models}
\label{subsection:Network_models}

\begin{figure}[ht]
\centering
\includegraphics[width=0.95\linewidth]{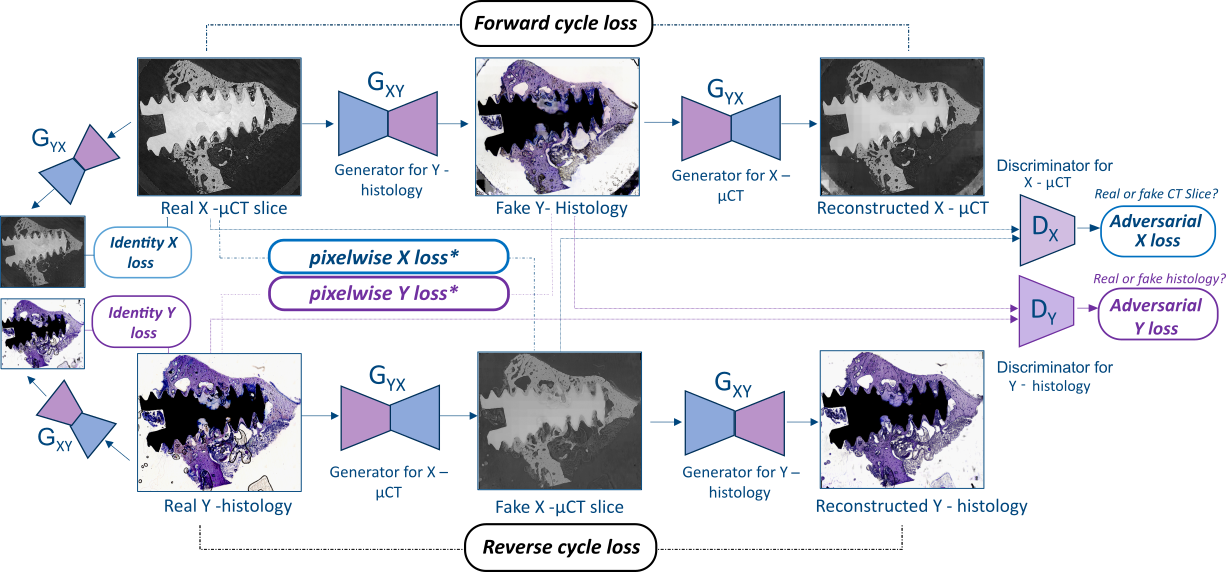}
\caption{The modified CycleGAN for paired data, with additional pixelwise supervision loss terms that directly penalise differences between the generator output and the target image for each domain. Not shown: an additional greyscale loss term applicable to the generated µCT data only.}
\label{fig:full-cycle-schematic}
\end{figure}

\subsubsection{CycleGAN and its adaptation for paired data}
\label{subsection:CycleGAN_adaptation}

Our chosen model is based on the CycleGAN framework and adapted for paired data. We worked with a PyTorch implementation of the original CycleGAN model \citep{zhu_unpaired_2017} as developed by \citet{linder-noren_pytorch_2019}. A schematic of our modified model is illustrated in Figure \ref{fig:full-cycle-schematic}.

The base model employs two generator networks: \( G_{XY} \), which learns to translate images from the source domain \( X \) (comprising µCT scans) to the target domain \( Y \) (histology images), and \( G_{YX} \), which performs the inverse mapping. Each generator is trained adversarially against a corresponding discriminator (\( D_Y \) and \( D_X \)), which learns to distinguish real from generated images.

Since in the original case no direct correspondence was assumed to exist between images in the two domains, the cycle-consistency constraint is introduced to ensure that an input image can be approximately recovered after translation and back-translation. Specifically, given an image \( x \in X \), the composition \( G_{YX}(G_{XY}(x)) \) should closely match the original \( x \). This acts as a form of self-supervision, allowing the network to learn meaningful mappings even without paired data. The cycle-consistency loss is defined as:
\begin{align}
\label{eqn:cycle_loss}
\mathcal{L}_{cyc}(G_{XY}, G_{YX}) &= \mathbb{E}_{x} \left[ \left\| G_{YX}(G_{XY}(x)) - x \right\|_1 \right] + \mathbb{E}_{y} \left[ \left\| G_{XY}(G_{YX}(y)) - y \right\|_1 \right]
\end{align}
whereby an \( \ell_1 \) (or mean absolute error) loss is applied between the original images and the reconstructed images, and $\mathbb{E}_{x}$ and $\mathbb{E}_{y}$ denote expectations over the respective data distributions.

For the adversarial loss: we use a mean squared error, or $\ell_2$, loss function between discriminator predictions and real/fake labels. The generator adversarial loss is given by:
\begin{equation}
\label{eqn:adv_loss}
\mathcal{L}_{\text{GAN}}(G_{XY}, G_{YX}) = \mathbb{E}_{y} \left[ (D_X(G_{YX}(y)) - 1)^2 \right]+\mathbb{E}_{x} \left[ (D_Y(G_{XY}(x)) - 1)^2 \right].
\end{equation}

An additional identity loss encourages preservation of colour and content when mapping images already in the target domain, of the form:
\begin{equation}
\label{eqn:identity_loss}
\mathcal{L}_{\text{id}}(G_{XY}, G_{YX}) = 
\mathbb{E}_{y} \left[ \lVert G_{XY}(y) - y \rVert_1 \right] + 
\mathbb{E}_{x} \left[ \lVert G_{YX}(x) - x \rVert_1 \right].
\end{equation}

Equations \ref{eqn:cycle_loss},\ref{eqn:adv_loss}, and \ref{eqn:identity_loss} represent the three loss functions commonly comprising the standard CycleGAN model. The total generative loss function for the original CycleGAN model combines these three components, weighted by hyperparameters \( \lambda_{\text{cyc}} \) and \( \lambda_{\text{id}} \):
\begin{equation}
\label{eqn:standardCycleGAN}
\mathcal{L}_{\text{total(unpaired)}} = 
\mathcal{L}_{\text{GAN}} + 
\lambda_{\text{cyc}} \cdot \mathcal{L}_{\text{cyc}} + 
\lambda_{\text{id}} \cdot \mathcal{L}_{\text{id}}.
\end{equation}

For our \textbf{modified CycleGAN} we adapt this model to paired data, whereby each image \( x \in X \) is assumed to have a corresponding ground truth image \( y \in Y \), and vice versa. To take advantage of this, we introduce a `pixelwise supervision loss' that directly penalises differences between the generator output \( \hat{y} = G_{XY}(x) \) and the target image \( y \), as well as \( \hat{x} = G_{YX}(y) \) and the target image \( x \). This is formulated as a combined \( \ell_1 \) loss:
\begin{equation}
\label{eqn:pixelwise_supervision_loss}
\mathcal{L}_{\text{px}}(G_{XY},G_{YX}) = 
\mathbb{E}_{(x, y)} \left[ \lVert G_{XY}(x) - y \rVert_1 \right]+\mathbb{E}_{(x, y)} \left[ \lVert G_{YX}(y) - x \rVert_1 \right].
\end{equation}

Lastly, we introduce a loss term which helps ensure that the generated µCT image is purely greyscale:
\begin{equation}
\label{eqn:greyscale_loss}
\mathcal{L}_{\text{gs}}(G_{YX}) = 
\mathbb{E}_{y} \left[
\lVert G^r_{YX}(y) - G^g_{YX}(y) \rVert_1 + 
\lVert G^r_{YX}(y) - G^b_{YX}(y) \rVert_1 + 
\lVert G^g_{YX}(y) - G^b_{YX}(y) \rVert_1
\right],
\end{equation}
where $G^r_{YX}(y)$,  $G^g_{YX}(y)$,  $G^b_{YX}(y)$ are the red, blue and green channels of the generated µCT image. (To present as a greyscale image, all 3 channel values should be equivalent.)

The total generative loss for our paired datasets combines all components, weighted by hyperparameters \( \lambda_{\text{cyc}} \) and \( \lambda_{\text{id}} \), \( \lambda_{\text{px}} \) and \( \lambda_{\text{gs}} \):
\begin{equation}
\label{eqn:modifiedCycleGAN}
\mathcal{L}_{\text{total(paired)}} = 
\mathcal{L}_{\text{GAN}} + 
\lambda_{\text{cyc}} \cdot \mathcal{L}_{\text{cyc}} + 
\lambda_{\text{id}} \cdot \mathcal{L}_{\text{id}} +
\lambda_{\text{px}} \cdot \mathcal{L}_{\text{px}} + 
\lambda_{\text{gs}} \cdot \mathcal{L}_{\text{gs}} 
\end{equation}

In this study, we evaluate the performance of the CycleGAN network as applied to paired data, using both the standard total generator loss (Equation \ref{eqn:standardCycleGAN}) and our modified version (Equation \ref{eqn:modifiedCycleGAN}). 

\subsubsection{Pix2Pix}

As a further baseline comparison, we test our modified CycleGAN for paired data against the classic Pix2Pix model \citep{isola_image--image_2018}, which is a conditional generative adversarial network (cGAN).  In this framework, the generator \( G \) learns to map an input image \( x \) (here a µCT slice) to a corresponding target image \( y \) (histological section), i.e., \( G(x) \approx y \). The discriminator \( D \) is explicitly conditioned on the input \( x \) and is trained to distinguish between real image pairs \( (x, y) \) and fake pairs \( (x, G(x)) \).

The conditional adversarial loss between generator and discriminator can again be written as an $\ell_2$ loss:
\begin{equation}
\label{eqn:pix2pix_adv_loss}
\mathcal{L}_{\text{cGAN}}(G) = \mathbb{E}_{x} \left[ (D(x,G(x)) - 1)^2 \right].
\end{equation}

This adversarial loss is balanced in the Pix2Pix model with a pixelwise loss commonly referred to the $\ell_1$ loss:
\begin{equation}
\label{eqn:pix2pix_pixelwise_loss}
\mathcal{L}_{\text{\( \ell_1 \)}}(G) = 
\mathbb{E}_{(x, y)} \left[ \lVert G{(x)} - y \rVert_1 \right].
\end{equation}
The total generative loss for Pix2pix then combines them with the weighted parameter \( \lambda_{\text{\( \ell_1 \)}} \):
\begin{equation}
\label{eqn:pix2pix_total_loss}
\mathcal{L}_{total}=\mathcal{L}_{cGAN}+\lambda_{\text{\( \ell_1 \)}}\mathcal{L}_{\text{\( \ell_1 \)}}
\end{equation}

\subsubsection{Model architectures and hyperparameters}
\label{subsub:architecture}

The base model architecture for both the CycleGAN model variations and Pix2Pix models tested follow the Pytorch implementations of \citet{linder-noren_pytorch_2019}. All models were initialised from scratch without pre-training before introducing our input data.

For the CycleGAN model variations: the generator network is based on a ResNet backbone with 9 residual blocks and a starting feature width of 64. The discriminator follows an unconditional PatchGAN design with a 70 $\times$ 70 pixel receptive field. To stabilise adversarial training, an image buffer was used to store previously generated samples for the discriminator loss, following the technique described in \citet{zhu_unpaired_2017}. Training was conducted using the Adam optimiser with a learning rate of $l_r = 0.0002$, and momentum parameters $\beta_1 = 0.5$ and $\beta_2 = 0.999$. For our modified CycleGAN loss model for paired data given in equation \ref{eqn:modifiedCycleGAN}, we achieved good results with loss factors $\lambda_{cyc} = 6$, $\lambda_{id}  = 3$, $\lambda_{px}=6$ and $\lambda_{gs}  = 1$. These factors are approximately proportional to those relative to the GAN loss as optimised in the original study by \citet{zhu_unpaired_2017}. The latter two factors are effectively ablated for the comparison with the standard CycleGAN model given by equation \ref{eqn:standardCycleGAN}.
Sample correspondence masks (see Subsection \ref{subsection:sample_correspondence}) were applied as binary multipliers for all $ \ell_{1} $ loss terms. 

For the Pix2Pix model, the standard architecture as introduced by \citet{isola_image--image_2018} consists of a U-Net generator and a conditional PatchGAN discriminator. Here the total loss model follows equation \ref{eqn:pix2pix_total_loss}, where $\lambda_{\text{\( \ell_1 \)}}=20$ after tuning. We also applied the sample correspondence mask to the $ \ell_{1} $ pixel loss term. All other architectural and training settings were kept consistent with those used in the CycleGAN model above. 

In each case models were trained on 256 × 256 pixel patches randomly sampled from 40 WSI pairs, with 100 patches per pair per epoch, a batch size of 4, and a total of 500 epochs. At each epoch, new randomised data augmentations were applied on-the-fly during patch extraction. During training, model states corresponding to the minima of each individual generator loss component were saved as checkpoints for later evaluation. Each training fold required approximately 100 hours on an NVIDIA Tesla V100 GPU for the CycleGAN models versus 40 hours for Pix2Pix, across the five-fold cross-validation scheme. 

\subsection{Data-augmentation}
\label{subsection:data_augmentation}
The utility of data-augmentation in achieving learning invariance has been well demonstrated \citep{dosovitskiy_discriminative_2014,navab_u-net_2015} and is particularly beneficial in medical imaging applications, where obtaining large numbers of paired datasets is often a challenge. By combining the randomised patch-based sampling with further data augmentation at each epoch we were able to simulate a substantially larger paired dataset than the initial set of WSI image pairs, thereby helping to prevent overfitting.
For each training example in both histology and µCT images, a random crop position was determined for the sampled patch. We also incorporated rescaling by 90-110\% (whilst maintaining patch array size), horizontal and vertical flipping, and a separate contrast and brightness jitter (up to 10\%) for both the µCT and histology patches. For the latter, the chromaticity or perceived colour is not varied (RGB ratios are preserved).

\subsection{Sample correspondence}
\label{subsection:sample_correspondence}
Sample correspondence masks were applied to the $\ell_{1}$ loss terms during training in order to exclude regions of the image pairs where there was known to be a mismatch between µCT and histology. These could be either in the imaging field of view or physical differences in the sample between tomographic and histological measurement arising from the sectioning process. The maps were created using a convex hull mask of the rigid bone tissue+screw sample, calculated via intensity-based segmentation. This method includes any soft tissue which is surrounded by the rigid bone material which is assumed to remain part of the sectioning, as well as a small amount of background pixels surrounding the bone. This effectively excludes from the training a majority of extraneous sample material (mainly soft tissue) in the µCT volume which did not end up in the histology section, as well as the walls of the sample holder which contained the bone-implant block. These walls can be seen in many of the CT slices. We generated convex hull masks for both CT and histology and took the minimum union, combined with a 3rd/4th mask generated with semi-manual selection for regions of any missing other information not accurately defined by the convex hull mask (e.g., histology images with missing sections of bone due to irregular sectioning). In general we did not attempt to mask out any small imperfections in the histology images, such as cracks in the optical slides, extra stain droplets, etc.
The masks were also subsequently applied to our showcased results and during their comparative analysis, whereby any pixels outside of the sample correspondence mask were replaced with a value equivalent to a mean background value.  

\subsection{Whole slide image outputs}
\label{subsubsection:wsi_outputs}

Within the field of virtual staining (with visible light), most methodological studies are constrained to patch-level analysis due to the gigapixel scale of WSIs at their original resolution \citep{liu_generating_2024}. Since we work with downscaled WSIs at the resolution of the input µCT data, we do not have the same memory-related constraints. Following the patch-based training of our model, we re-tiled the model outputs to obtain overlapping patch-based inference of whole slide images of the order 1k $\times$ 1k pixels. Analysis of model performance across the WSI field of view has several advantages over a patch-level analysis. This approach provides a more comprehensive assessment of output quality, as localised patches may appear visually convincing in isolation (passing the visual Turing test) but fail to reflect broader inconsistencies. In contrast, overlapping patch-based WSIs may reveal global artifacts such as tiling borders or inconsistent transitions that arise in cases of model instability. Although direct WSI inference is also possible given network scalability,  we nevertheless retain patch-based inference for consistency with training and to preserve the same spatial frequency sampling.

\subsection{Comparative metrics}
\label{subsection:Comparative_methods_metrics}

In addition to a visual assessment of each model's performance, we also applied the following complementary metrics for a comprehensive quantitative comparison: Structural Similarity Index Measure (SSIM), Peak Signal-to-Noise Ratio (PSNR), and Learned Perceptual Image Patch Similarity (LPIPS). These metrics capture different aspects of image similarity, including structural coherence, pixel-wise differences and perceptual fidelity.

We applied the metrics through patch-based (256 $\times$ 256 pixels) sampling of the inferred WSI histology images with reference to their input histology counterparts. Background patches were excluded from analysis according to the sample correspondence mask (defined by a 50$\%$ overlap or greater). 

The Structural Similarity Index Measure, or SSIM \citep{wang_image_2004} quantifies perceptual similarity between two images based on luminance, contrast, and structural components. It is defined as:
\begin{equation}
\text{SSIM}(x, y) = \frac{(2 \mu_x \mu_y + C_1)(2 \sigma_{xy} + C_2)}{(\mu_x^2 + \mu_y^2 + C_1)(\sigma_x^2 + \sigma_y^2 + C_2)},
\end{equation}
where $\mu_x$ and $\mu_y$ are the mean intensities, $\sigma_x^2$ and $\sigma_y^2$ are the variances, $\sigma_{xy}$ is the covariance, and $C_1$, $C_2$ are small stability constants. We used the SSIM implementation by \citet{van_der_walt_scikit-image_2014}.

The Peak Signal-to-Noise Ratio (PSNR) is a traditional metric for quantifying absolute error of the reconstruction quality (between real and generated images) on a logarithmic scale:
\begin{equation}
\text{PSNR}(x, y) = 10 \cdot \log_{10} \left( \frac{L^2}{\text{MSE}(x, y)} \right),
\end{equation}
where $L$ is the maximum possible pixel value (i.e., 255 for 8-bit images), and $\text{MSE}(x, y)$ is the mean squared error between the images:
\begin{equation}
\text{MSE}(x, y) = \frac{1}{N} \sum_{i=1}^{N} \left( x_i - y_i \right)^2,
\end{equation}
where $N$ is the number of pixels. 

The Learned Perceptual Image Patch Similarity or LPIPS \citep{zhang_unreasonable_2018} measures perceptual similarity by comparing deep features extracted from a pretrained neural network. Unlike SSIM or PSNR, LPIPS captures high-level perceptual and colour differences, making it well-suited for evaluating visual fidelity in generative tasks in the RGB space. Given feature maps $f_l(x)$ and $f_l(y)$ at layer $l$, the LPIPS distance is:
\begin{equation}
\text{LPIPS}(x, y) = \sum_l \frac{1}{H_l W_l} \sum_{h,w} w_l \left\| \hat{f}_l^x(h,w) - \hat{f}_l^y(h,w) \right\|_2^2,
\end{equation}
where $\hat{f}_l$ denotes channel-wise normalized feature maps and $w_l$ are learned weights. We used the LPIPS implementation by \citet{detlefsen_torchmetrics_2022} with the SqueezeNet network type.

\subsection{3D testing}
\label{subsection:3D_testing}

Once our chosen network model was trained, we applied the forward generative model to a stack of input SR-µCT slices for validation in 3D. These were untransformed slices, although one training pair was also produced from this dataset through the co-registration process outlined in Subsection \ref{subsection:Registration}. A 3D gaussian filter with a sigma of 1 pixel was first applied to the CT slice stack, in order to simulate the resolution reduction which occurs through interpolation steps of the co-registration process. The output stained slices were generated from the trained forward model independently as single slices. For 3D visualisation these slices were loaded into Avizo3D 2024.2 (Thermo Fisher Scientific, Berlin, Germany), and the mean of the RGB channels was computed as the alpha (A) channel. This enabled a volume rendering with direct RGBA mapping. 

\section{Results and Discussion}
\label{section:Results_and_discussion}

\begin{figure}
\centering
\begin{subfigure}{0.5\textwidth}
   \includegraphics[width=\textwidth]{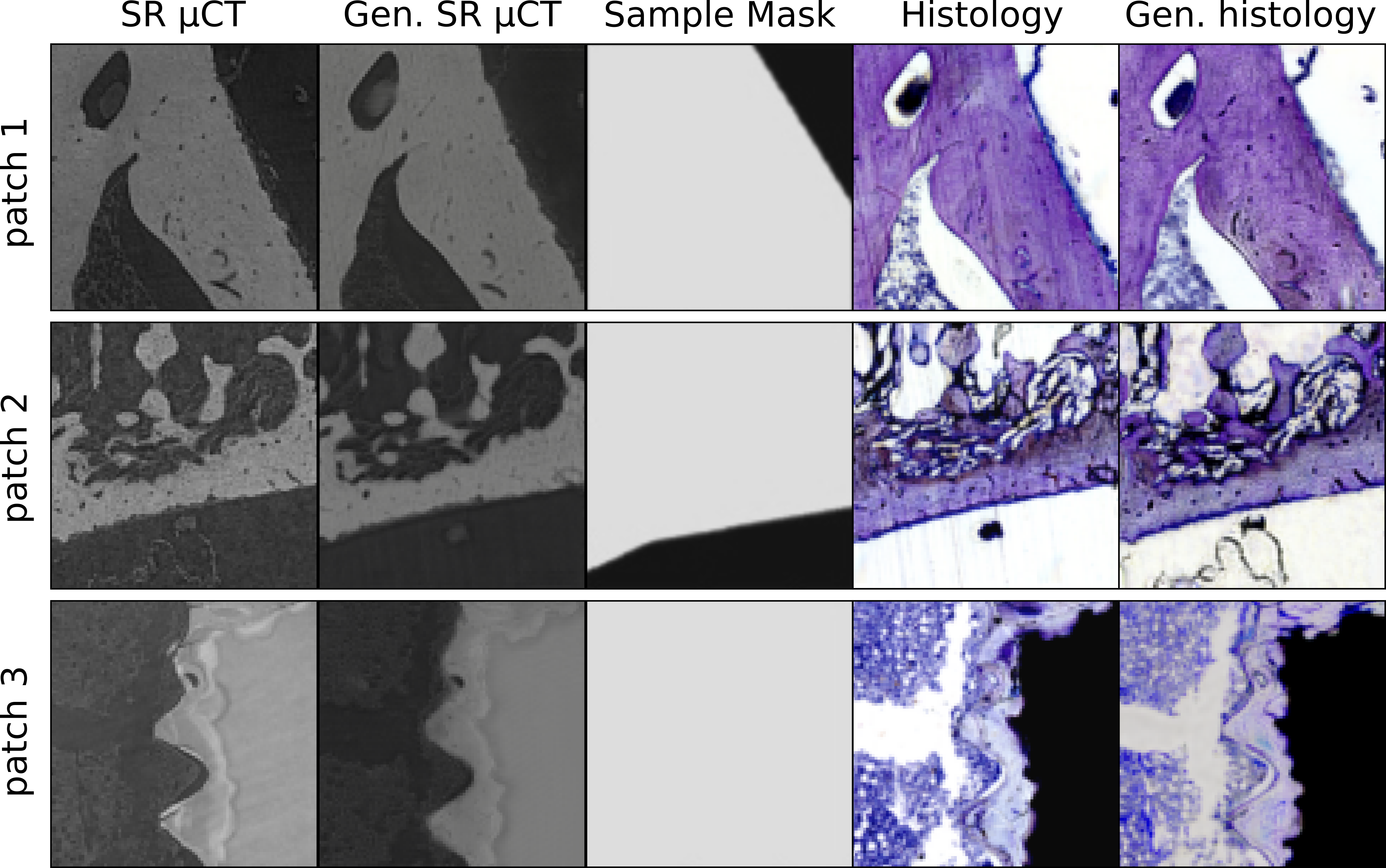}
    \caption{}
    \label{fig:patch_example}
\end{subfigure}
\hfill
\begin{subfigure}{0.95\textwidth}
    \includegraphics[width=\textwidth]{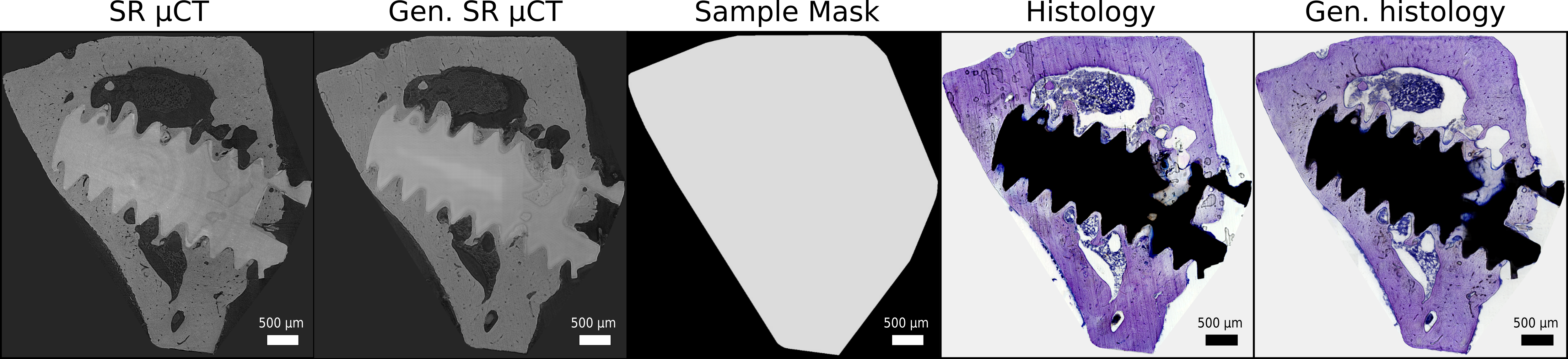}
    \caption{}
   \label{fig:wsi_example}
\end{subfigure}   
\caption{Examples of our modified CycleGAN model training results: (a) direct training patch input/outputs (256 x 256 pixels); and (b) example WSI result generated through overlapping patch-based inference, from 1 WSI sample pair included in training (958 $\times$ 1000 pixels). The displayed µCT and histology images have been masked (replacing all background pixels with the mean value) with the included sample correspondence mask.}
\label{fig:patch_vs_wsi}
\end{figure}

An example of the direct training patch results of our modified CycleGAN model is shown in Figure \ref{fig:patch_example}. For each 256 $\times$ 256 pixel patch region, there is the input real SR-µCT patch, and output generated histology patch, as well as the input real histology patch and corresponding output generated µCT patch. In addition, the sample correspondence map patch is also shown in the central panel.  At the patch level, the similarity between input and generated images appears very high. However, it is possible to see that the patches are not perfectly aligned down to every pixel. This is partly due to the initial registration, which was performed solely on a global scale of the WSI and not subsequently re-applied to each individual patch, as well as to differences in the sample slice pair which arise from the histological sectioning process. In particular, the soft bone tissues (as shown in blue to the left of patch 3 of Figure \ref{fig:patch_example}) were observed to shift and these regions are not co-registered as accurately as the rigid bone structures. In Figure \ref{fig:wsi_example}, a representative WSI training result is shown, as generated through overlapping patch-based inference (stepping the 256 $\times$ 256 pixel patch across with a step size of 64 pixels). The sample correspondence mask (shown in the middle panel) has been applied after image generation whereby pixels exterior to the sample are set to the mean background value. Examples of the inferred WSI training, validation and test results displayed without the mask may be seen in the Supplementary section (see Figures \ref{fig:sup_train_results}, \ref{fig:sup_val_results}, and \ref{fig:sup_test_results}). Behind the mask, the effect of overlapping patches can sometimes be observed in background regions of the generated images in the form of tiling artefacts of varying intensity steps, reflecting uncertainty in the model prediction due to exclusion from the training process (via the sample correspondence mask). The step size may be reduced to smooth this effect, however generally in well-trained image regions there are no such artefacts. Note that for the remainder of this manuscript, generated image results will be presented as masked overlapping patch-based inferred WSI images, or a region of interest thereof, and not the original patches directly used as model input/output.

\begin{figure}
\centering
\begin{subfigure}{0.98\textwidth}
   \includegraphics[width=\textwidth]{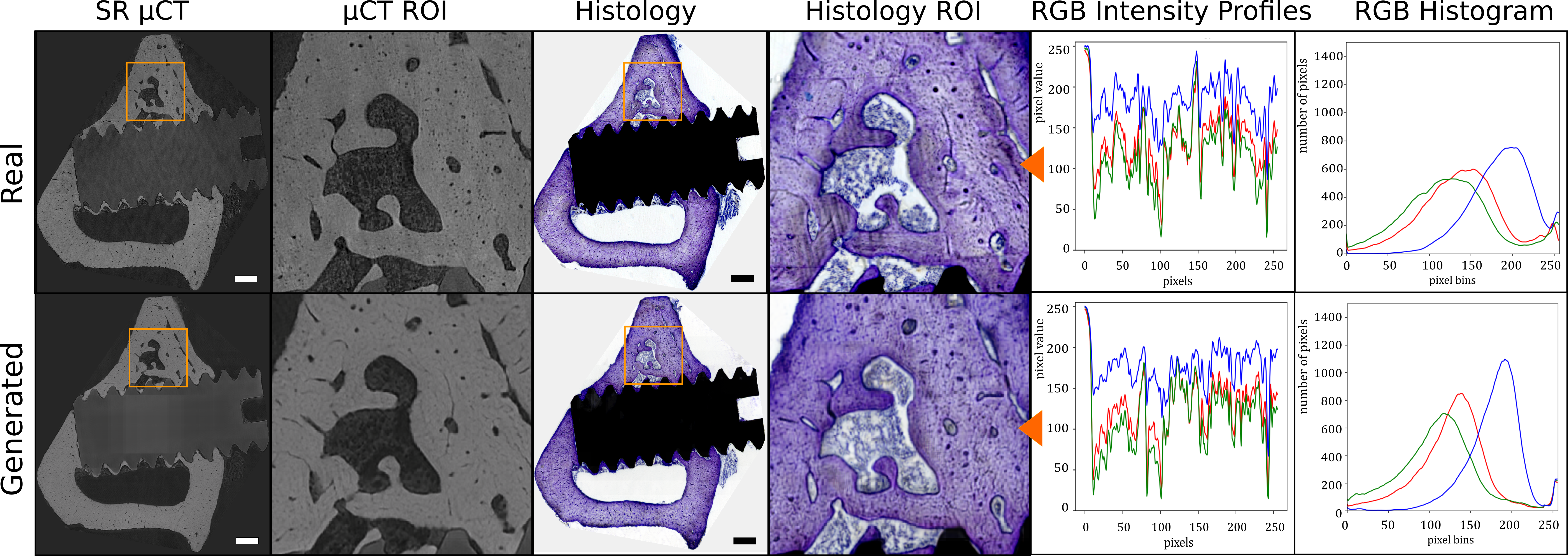}
    \caption{training}
    \label{fig:results_training}
\end{subfigure}
\hfill
\centering
\begin{subfigure}{0.98\textwidth}
   \includegraphics[width=\textwidth]{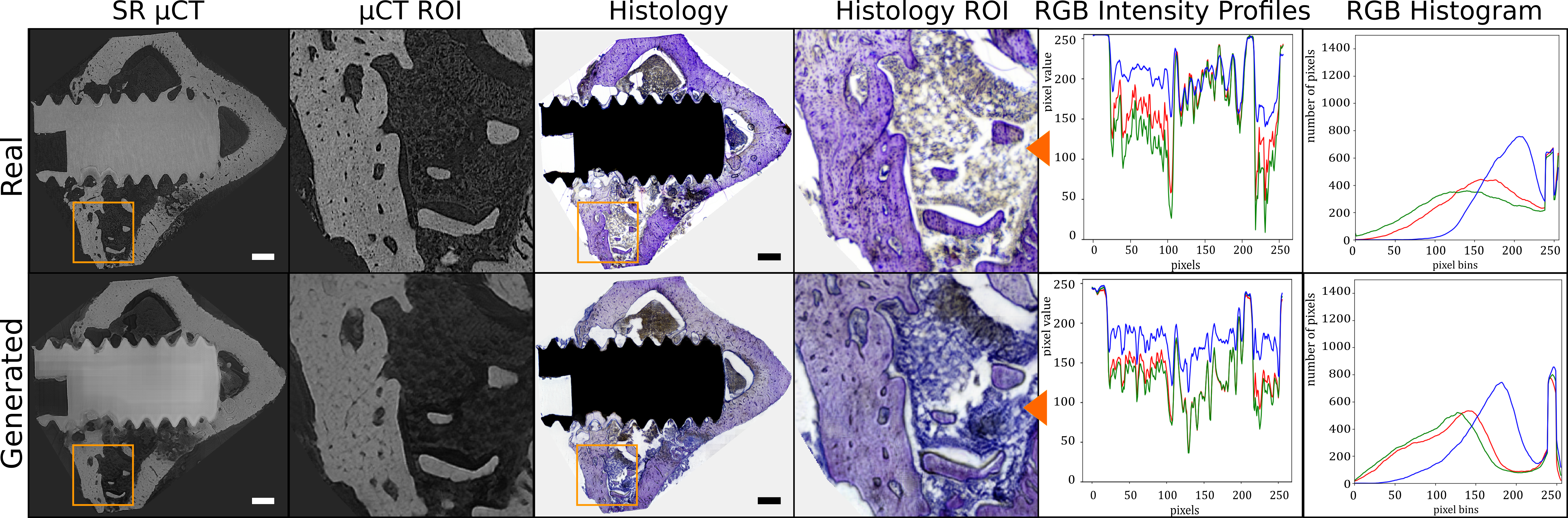}
    \caption{validation}
    \label{fig:results_validation}
\end{subfigure}
\hfill
\begin{subfigure}{0.98\textwidth}
    \includegraphics[width=\textwidth]{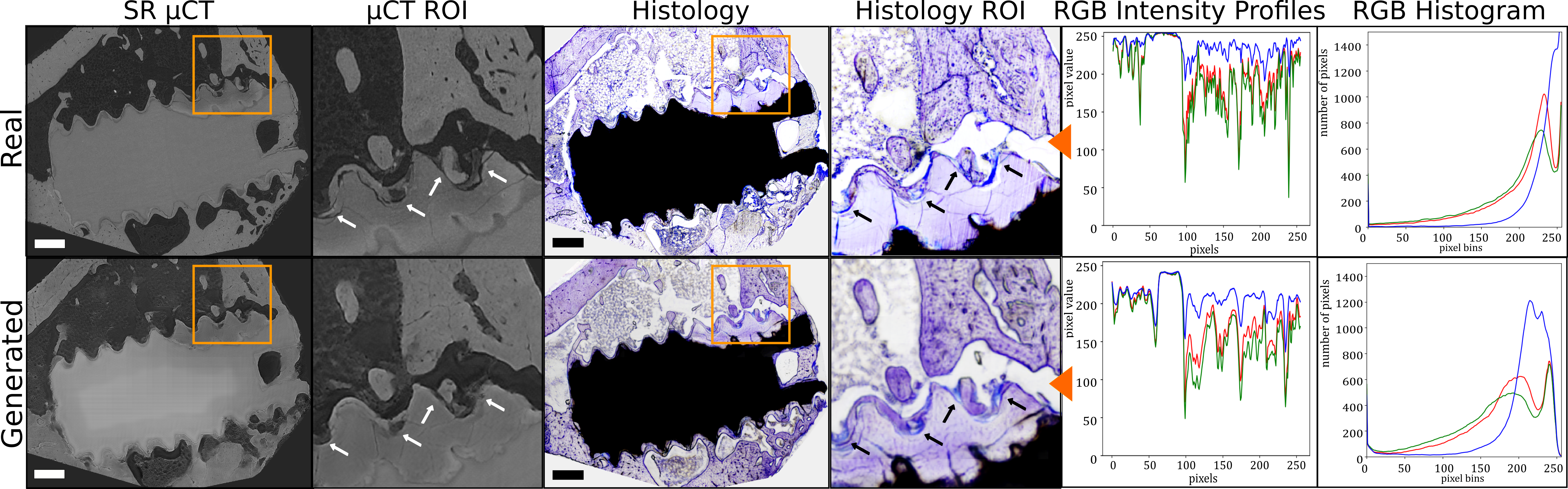}
    \caption{testing}
   \label{fig:results_test}
\end{subfigure}          
\caption{Example modified CycleGAN model results (with WSI output) for (a): training, (b) validation and (c) testing. Includes a 256 $\times$ 256 pixels region of interest (ROI) with accompanying RGB histograms and intensity profiles of real vs generated examples. Each profile is acquired horizontally across the midpoint of the ROI as indicated by the orange arrow. The scalebar in each WSI represents 500 µm. In (c), isolated regions of new bone growth are indicated with arrows in white (µCT) and black (histology), for both the input data and generated histology.} 
\label{fig:results_validation_testing_profiles}
\end{figure}

Representative results of our modified CycleGAN model from training, validation and testing datasets are demonstrated in Figure \ref{fig:results_validation_testing_profiles}. Overall there is a good apparent level of agreement between real and generated WSIs in both domains for each dataset, although as to be expected, there is a notable drop in perceived match accuracy when comparing against the training results. In particular the generated WSI histology training results appear excellently matched. 

In general, the forward model (SR-µCT to predicted histology) results of the CycleGAN are more likely to pass a visual Turing test than the reverse model (histology to predicted SR-µCT). Within the generated µCT slices, tiling artefacts (as mentioned above) are present in the majority of regions containing the residual screw alloy (represented in black pixels within the corresponding histology). Unlike the background regions, these screw regions were not excluded from the training process by the sample correspondence masks.  However, the uncertainty in their prediction lies in the possibility of three different material components (PEEK, Mg or Ti) within the sample pool which are characterised by three different linear attenuation coefficients for SR-µCT (respectively: less than, approximately equal to, or greater than the value for dense bone tissue). Each of these are matched to the same black values of the histology input. In our preliminary studies previously reported in \citet{irvine_multi-modal_2024}, we initially incorporated a fourth channel (in addition to R,G, and B channels) as model input in which the known screw material could be indicated as a single value. However, this extra parameter was eventually found to contribute to a significant blurring effect over the whole generated image and we have since abandoned this in our current model.
Further contributing to their synthetic appearance, the generated CT slices tend to exhibit reduced high-frequency detail compared to their real counterparts, and lack the shot noise characteristic of real X-ray images.

Other noted discrepancies between real and generated images include the many characteristic features of the real histology sample images which have not been replicated in the generated histology. These include cracks in bone, saw marks (i.e., one-directional striations), histology slide fractures, and stain droplet spills. The same histology features were mostly reproduced in greyscale format within the generated CT slices, also distinguishing them from their real CT counterpart. All of these features were generally included in the training (unless located significantly external to the bone sample in which case they were masked out), however, they were not able to be `learned' as features due to a lack of correspondence. These features are not intrinsically related to the sample but are formed as part of the histological sample preparation process, and are absent from the CT acquisition. This highlights the distinction between accuracy, referring to faithful structural representation, and realism, which encompasses visual artefacts that enhance stylistic plausibility. The omission of such features reflects the model’s preference for preserving structural accuracy over superficial realism.

The colour, contrast and resolution fidelity of our paired CycleGAN output results are demonstrated in more detail with the selected ROIs and corresponding RGB histogram plots and intensity profiles derived from the histology ROIs, within Figure \ref{fig:results_validation_testing_profiles}. As expected, the training results are a superior match to the validation and test results, most notably with respect to the colour fidelity of the generated histology, but also in the reproduction of the finest features. In the training data (Figure \ref{fig:results_training}), both real µCT and histology ROIs exhibit a large number of medium and small-sized pores visible down to only a few pixels in size within the dense bone structure, which correspond to the blood vessels and lacunae respectively. These pores are well reproduced in the generated histology, demonstrating the high level of detail able to be supported by the forward model. Accordingly, the histograms and profiles are very similar although non-identical.
In the validation data set (Figure \ref{fig:results_validation}), the soft tissue in the generated histology is notably of a different colour in certain areas to the real image (with respect to blue and brown), as well as differing in intensity, although the variations could be considered realistic as they reflect the varying colours and intensities of soft tissues observed elsewhere in the histology dataset. Such differences are reflected in the histogram, and in the RGB ratios at levels attributable to the soft tissue pixels. The lacunae and other fine features are also less visible in comparison to the training example.
In the test example (Figure \ref{fig:results_test}), discrepancies between real and generated histologies in the colour of both dense bone and soft tissue may also be observed. Conversely to the validation ROI above, the generated soft tissue here is skewed more towards brown than the blue of the real soft tissue. Additionally, we note that our test input µCT data is of lower resolution than the typical 50 training and validation CT examples. As a result, the generated histology also contains fewer fine features, which is visible in broadened histogram peaks and smoother line profile. However, the loss of detail is not as pronounced as one might expect based solely on the input CT resolution, particularly in the dense bone regions. In fact, while a comprehensive spatial frequency model analysis is beyond the scope of this study, preliminary findings (see Supplementary Section \ref{subsection:sup_resolution}) suggest that the GAN-generated histology typically gains a spatial resolution that is lower than real histology but higher than the input µCT. This places the output resolution between the two modalities and supports the qualitative observation that fine structural details may be partially preserved, despite the lower-resolution input.

As a final qualitative test, the selected testing ROI in Figure \ref{fig:results_test} provides a valuable demonstration of the GAN generator’s predictive accuracy in capturing certain key features from the original X-ray histology study, namely the Mg alloy screw degradation layer and potential regions of new (woven) bone growth (see also Figure \ref{fig:comparisonCThisto} for reference).
Here we observe that the prediction for the degradation layer exhibits a boundary that is reasonably consistent with the real histology although a blurriness at the edge of the black residual screw suggests some uncertainty. Upon closer inspection within the degradation layer, we note that while in real histology images this area is characteristically devoid of bone cells \citep{kruger_high-resolution_2022}, inside the generated layer a faintly repeating cell-like structure is apparent. This suggests that some features may be falsely enhanced from image noise, and more representative data of degradation layers is ideally required for the model to train upon. Importantly, the generated model appears to have successfully predicted the presence of new bone, as indicated by the bright blue regions next to the degradation layer (annotated in the ROI by small black arrows) which were matched to areas of reduced density within the CT image (white arrows). 

Results of our modified CycleGAN model on the secondary H\&E-stained dataset are provided in Supplementary Section \ref{subsection:si_HEdescription}. While the dataset is too small for extensive interpretation, these results are consistent with our main findings and suggest that the model may generalise across multiple stains.

\subsection{Qualitative comparison of models}
\label{subsection:models}

Here we present a comparative analysis of our modified CycleGAN model (including extra $\ell_{1}$ pixelwise supervision and greyscale loss terms given in Equation \ref{eqn:modifiedCycleGAN}) as evaluated against Pix2Pix representing the baseline model for paired data, as well as the standard CycleGAN model, which also receives paired inputs but is trained using only the standard generator loss (Equation \ref{eqn:standardCycleGAN}). Whilst the first two models both performed in a predictable manner, the standard CycleGAN model was observed to perform poorly and did not converge to a stable equilibrium. As the only model without direct supervisory loss terms, it was unable to overcome the strong intensity mismatch between the CT and histology domains. The forward generator attempted to create histology images with a dark background and light bone structure, whereas the reverse generator tried to create CT images with a bright background and dark bone structure. This issue is consistent with challenges previously reported when applying CycleGAN to virtual staining tasks with unlabeled microscopy images which present an inverted contrast to the desired stained target images \citep{bai_deep_2023}. To mitigate this, we adopted a simple recommended workaround \citep{chen_deep-learning-assisted_2021, abraham_mode-mapping_2022} and performed a second set of tests of the standard CycleGAN model whereby the input CT images were first inverted. This adjustment led to some limited improvement in both output quality and training stability. The following results are thus included for each of four model variants which includes the standard CycleGAN with both original and contrast-inverted CT inputs. For all variants we focus exclusively on the forward output (i.e., the generated histology), as this represents the primary objective of the application. Additionally, because Pix2Pix is composed of only one generator, it does not produce a corresponding CT output.

\begin{figure}[ht]
\centering
\includegraphics[width=0.95\linewidth]{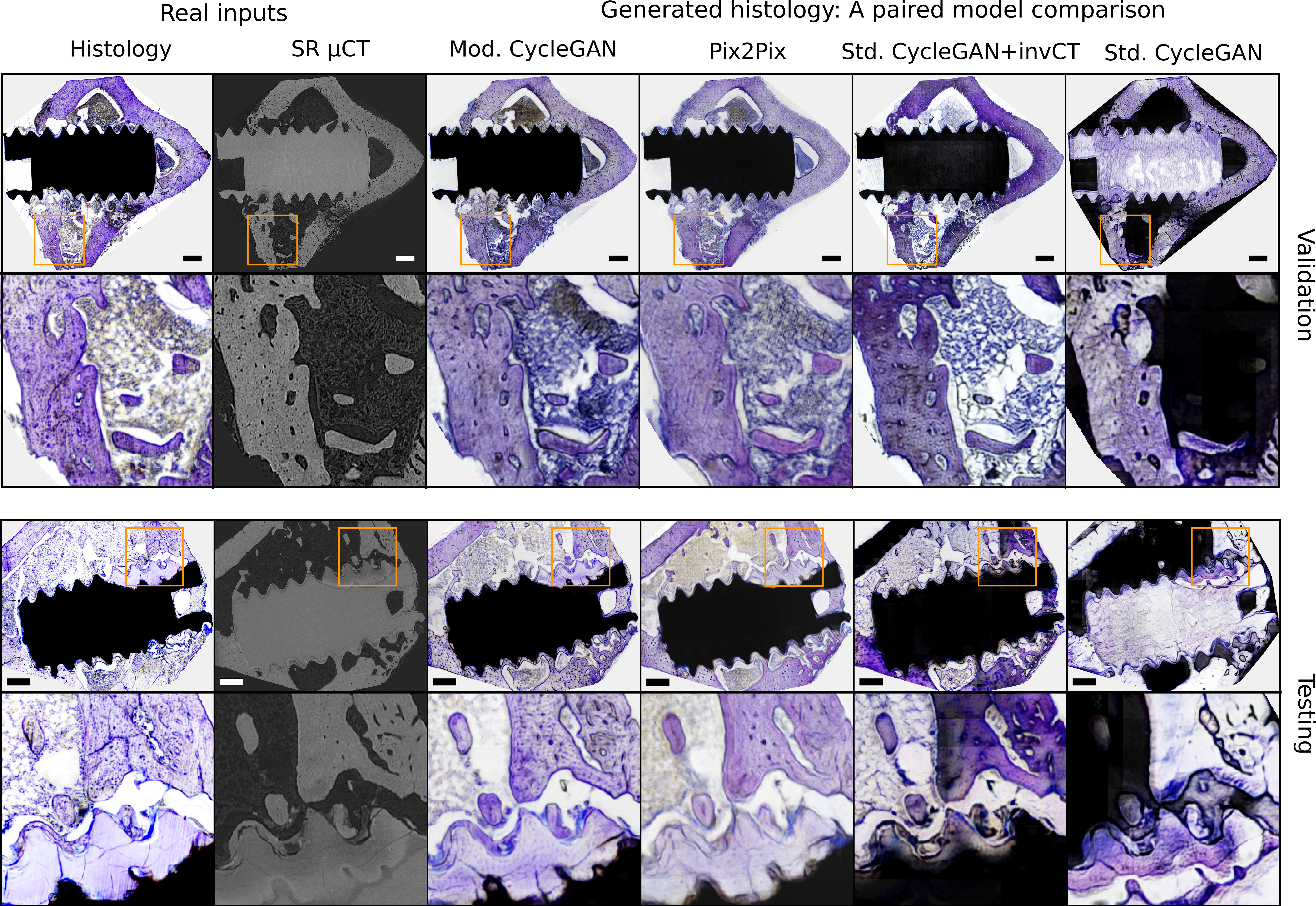}
\caption{Inference results from each of the paired model variants tested, with WSI and 256 pixel ROI examples from validation and test data sets. `Mod. CycleGAN' = the modified CycleGAN including the extra supervisory and greyscale loss terms, which is our chosen model. `Std. CycleGAN' = the Standard CycleGAN model, which was trained from the same paired data sets, but without the extra loss terms. `Std. CycleGAN + invCT' refers to a variant of the same model but trained and tested with an inverted CT input, in an attempt to combat the issue of intensity mismatch. The scalebar in each WSI represents 500 µm.}
\label{fig:results_model_comparison}
\end{figure}

In Figure \ref{fig:results_model_comparison} we qualitatively compare the inference results of all paired model variants with example validation and test outputs of a WSI with a $256 \times 256$ pixel ROI (the same ROI is reused from Figure \ref{fig:results_validation_testing_profiles}). Our model visibly outperforms the competing models. The Pix2Pix model was able to reproduce colour reasonably well; however, the resolution and texture of the generated histology images were noticeably inaccurate. During hyperparameter tuning an emphasis was placed on resolution, which reduced the weight of the $\ell_{1}$ loss relative to the adversarial term. The resulting increased GAN loss contribution yields some high-frequency artifacts and texture hallucination. The poor performance of Pix2Pix may also be attributed to its sensitivity to the imperfect alignment of paired data, particularly in finer features such as individual bone pores and in soft-tissue regions prone to deformation during histological sample preparation. As mentioned, the standard cycleGAN model (far right) performed very poorly without the additional supervisory loss terms, and using normal CT inputs. The model failed totally to replicate the appropriate histological colours with respect to bone material and background. After switching to an inverted CT input, we observed an increased performance with a more appropriate contrast range, but the model still had trouble with the black screw region of the histology images corresponding to a range of brightness values in the input CT (depending on implant material). This resulted in many regions of bone in the generated histology falsely predicted as black, as well as poor delineation of the degradation layer. Many pores and voids were also predicted with an inverted intensity. Despite a generally poor colour prediction, the standard CycleGAN without extra supervisory $\ell_{1}$ term was able to reproduce the fine features reasonably well, and in particular the soft tissue regions were more sharply defined than in our chosen model. 

Supplementary Section \ref{subsection:si_HEmodels} also includes a reduced model comparison on the H\&E-stained dataset, centred upon performance of the modified CycleGAN versus Pix2Pix. The results indicate that the modified CycleGAN maintains an advantage over the Pix2Pix baseline, further supporting its applicability across different staining protocols.

\subsection{Quantitative evaluation}
\label{subsection:quantitative}

\begin{table}
\resizebox{\textwidth}{!}{%
\begin{tabular}{l||c|c|c||c|c|c||c|c|c}
\hline
\multirow{2}{*}{Model} & \multicolumn{3}{c}{Metrics (training)} & \multicolumn{3}{c}{Metrics (validation)}& \multicolumn{3}{c}{Metrics (testing)}\\
& SSIM $\uparrow$ &LPIPS $\downarrow$& PSNR $\uparrow$ & SSIM $\uparrow$ & LPIPS $\downarrow$& PSNR $\uparrow$ & SSIM $\uparrow$ & LPIPS $\downarrow$& PSNR $\uparrow$ \\
\hline
Mod. CycleGAN (our model)& \textbf{0.63}& \textbf{0.12}& \textbf{17}& \textbf{0.60}& \textbf{0.16}& \textbf{15}& \textbf{0.59}& \textbf{0.16}& \textbf{15}\\
Pix2Pix& 0.24& 0.15& 16& 0.21& 0.17& 14& 0.22& 0.22& \textbf{15}\\
Std. CycleGAN + inv. CT input & 0.27& 0.25& 13& 0.31& 0.25& 13& 0.33& 0.25& 12\\
Std. CycleGAN& 0.15& 0.44& 5& 0.18& 0.43& 5& 0.20& 0.43& 5\\
%Standard CyleGAN  & 3466 & 0   & 1357    \\
\end{tabular}}
\caption{\label{table:metrics} Evaluation of models with median values of the metrics SSIM, LPIPS and PSNR for training (from 40 WSI samples), validation (10 WSI samples) and testing (3 WSI samples). The top values are highlighted with bold text.}
\end{table}

\begin{figure}[ht]
\centering
\includegraphics[width=0.9\linewidth]{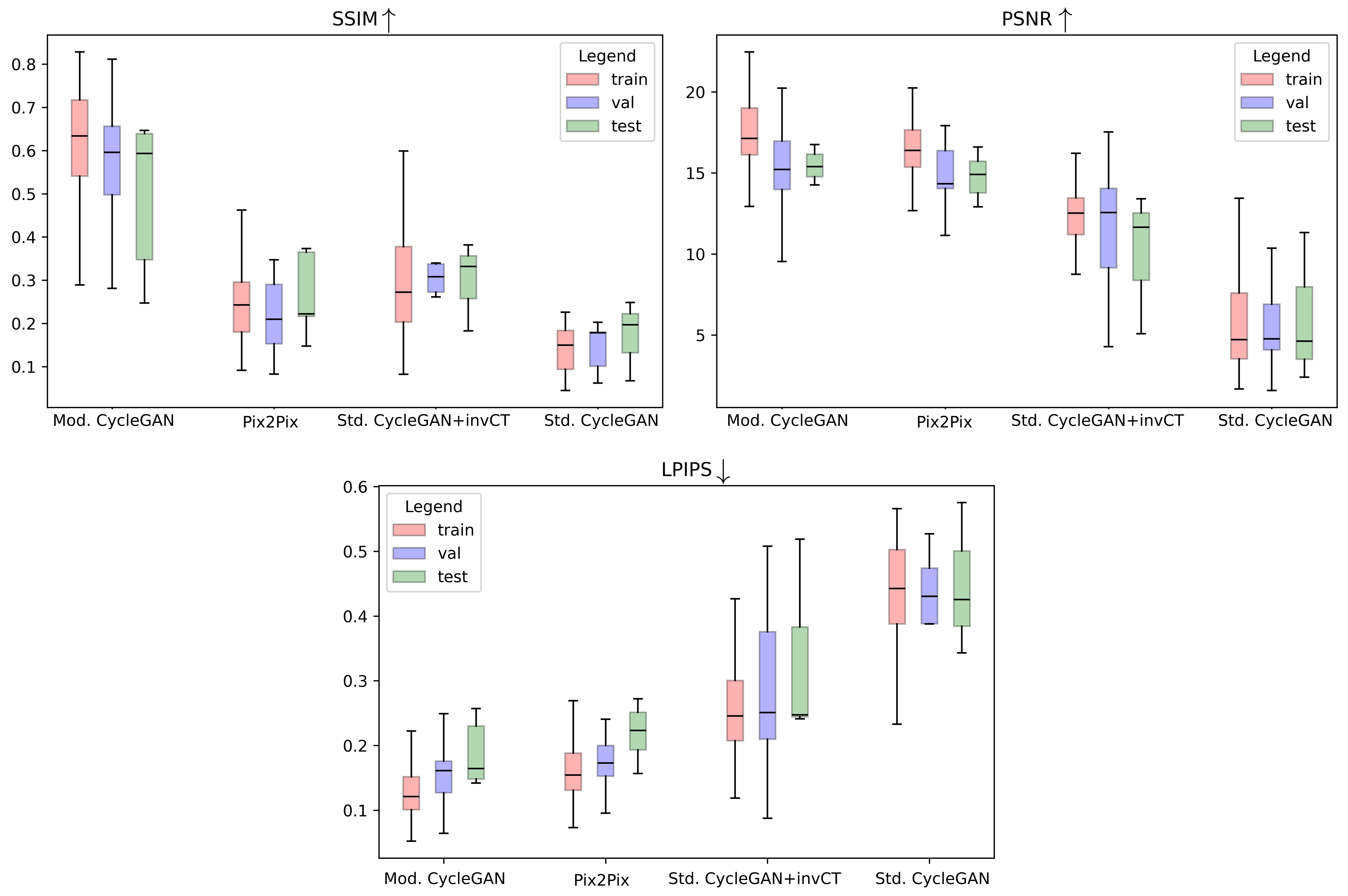}
\caption{Box plots for a comparison of the paired model variants tested, across the three metrics of Structural Similarity (SSIM) metric, Peak Signal-to-Noise Ratio (PSNR), and Learned Perceptual Image Patch Similarity (LPIPS).}
\label{fig:3metrics}
\end{figure}

Table \ref{table:metrics} presents the quantitative measurements of all paired model variants (our modified CycleGAN model with extra loss terms, compared to the standard Pix2Pix and standard CycleGAN models, the latter applied with both normal input µCT and inverted µCT input). Measurement values for each of the metrics: SSIM, PSNR and LPIPS are shown, for each of training, validation and testing samples. The values shown are the median calculated values from quite varied distributions due to the highly heterogeneous structures across the sampled WSIs. For this reason we have also included box plots of the distributions in Figure \ref{fig:3metrics} where it is possible to observe the range of values.  

When evaluated quantitatively, our model consistently outranks all other variants across the three metrics, with particularly strong performance in the structural similarity (SSIM) metric. Among the alternative model variants, the standard CycleGAN with uninverted µCT inputs clearly performed the worst overall. Between Pix2Pix and the standard CycleGAN with inverted µCT inputs, Pix2Pix achieved higher PSNR and lower LPIPS values, indicating lower reconstruction error and a good colour match. However, it performed significantly worse on SSIM, due to the observed high-frequency artefacts. Conversely, the standard CycleGAN with inverted µCT inputs exhibited better SSIM but poor LPIPS, reflecting higher structural fidelity but poor colour consistency.

It is worth noting that while the top scores remain relatively low compared to results from virtual stain transfer, for example, they are comparatively high for other cross-modality image translation tasks. All metric values are somewhat inflated due to the inclusion of screw regions, which comprise large areas of homogeneous black pixels, although these are considered characteristic of our model's application. As a baseline, when the same metrics were computed between the input µCT slices and their co-registered histology counterparts (i.e., without generation), the average SSIM, LPIPS, and PSNR were 0.15, 0.4, and 5, respectively. Thus, the standard CycleGAN outputs without inverted CT input are not substantially better matched to the real histology than the raw µCT inputs themselves.

\subsection{3D qualitative results from our model}
\label{subsection:3D-results}

\begin{figure}[ht]
\centering
\includegraphics[width=0.7\linewidth]{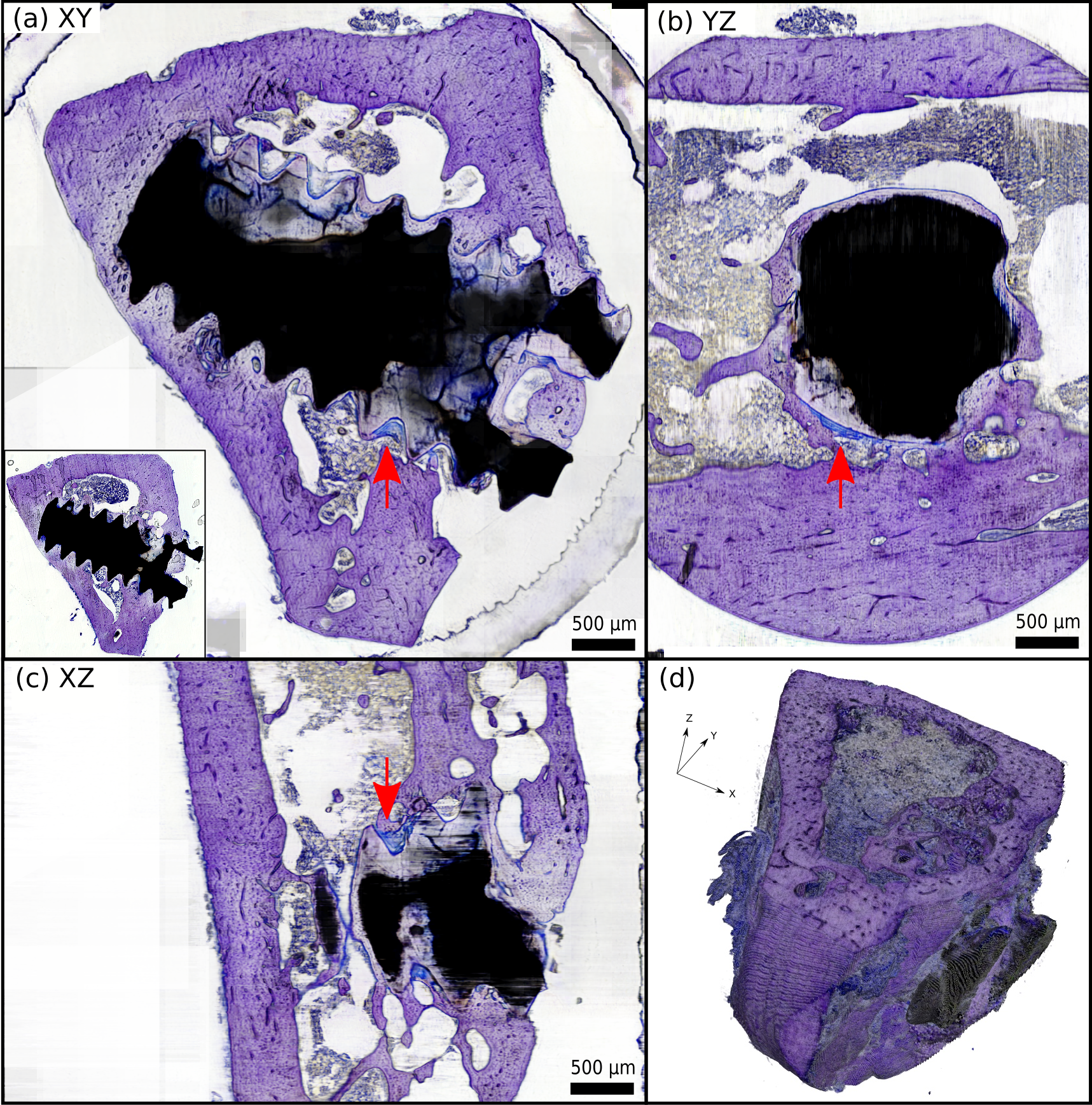}
\caption{Orthogonal views (a-c) and volume rendering with direct RGB  mapping (d) of a generated 3D X-ray virtual histology dataset featuring a Mg-based degradable bone-implant sample. A real histology image acquired from the same bone-implant sample is shown for reference in the bottom-left inset of (a). Red arrows point to a region of new bone growth adjacent to the degradation layer, distinctly coloured in a bright blue. }
\label{fig:3D_composite}
\end{figure}

Orthogonal slice views and a volume rendering of a virtually stained 3D X-ray histology dataset (featuring a Mg-based bone-implant) generated from our trained modified CycleGAN network are shown in Figure \ref{fig:3D_composite}. Overall, the generated volume displays strong visual consistency throughout. However, minor stripe artefacts are visible in the YZ and XZ planes (b–c), caused by slight intensity variations along the slice stack (Z) axis. These artefacts are more pronounced in regions containing the degradation layer, where the model shows increased slice-by-slice intensity variation due to uncertainty in predicting the precise interface between the residual alloy and the degradation material. Increasing the size of the training dataset would likely improve performance in this area. Alternatively, generating the volume using multiple orthogonal input planes and combining the outputs (similar to the multi‐axes prediction fusing approach outlined in \cite{baltruschat_scaling_2021}) should help produce a smoother and more coherent result. In this initial study, we opted for a simpler approach based on independently generated slices, as this best illustrates the model’s stability across the volume.

Notably, our model accurately predicts one of the key regions of interest within the original X-ray histology context, identifying multiple areas of new bone growth adjacent to the degradation layer of Mg-based implants. This newly formed bone, characterised by low mineralisation, is distinctly highlighted in bright blue within the 3D virtual staining. One such region is visible across all three orthogonal planes, as indicated by the red arrows. Upon closer re-inspection of the original input 3D CT dataset, the prediction was supported by correspondingly lower greyscale values in the same region, although these features were not immediately apparent. 

\subsection{Future outlook}
\label{subsection:future-outlook}

In this work, we showed that paired data is necessary for effective training in this specific application and domain, as a standard CycleGAN without supervisory loss terms performed poorly. However, the acquisition of paired multimodal imaging datasets in biomedical applications remains a significant challenge. Future work could explore hybrid model architectures that incorporate unpaired data, such as the approach by \citet{tripathy_learning_2018}, which initially trains on paired datasets before refining with unpaired samples.

The colour mismatches observed during testing, particularly in soft tissue regions, suggest the model may not yet fully generalise to the colour variability present in larger histology datasets. Addressing this could involve integrating more advanced colour normalisation techniques, despite the broader issue of colour standardisation in digital pathology.

While the results at the whole-slide image (WSI) level are strong and practical for 3D inference over volumes typical of X-ray µCT virtual histology, the resolution remains relatively low (greater than 5 µm) compared to conventional digital pathology WSIs. Future work will focus on evaluating model performance at higher resolutions, closer to original histology quality, which will likely require interpolating CT data to upscale the training input. This could effectively transform the model into a super-resolution generator and enable deeper investigation into its spatial frequency response. Data augmentation strategies that incorporate artificial noise and smoothing filters could additionally improve robustness.

A further extension could include the integration of segmentation labels such as used in 3D µCT analysis \citep{baltruschat_scaling_2021}, which would support both evaluation and conditional training in future 3D applications.

Finally, to enhance clinical relevance, collaboration with medical experts will be crucial to ensure biological staining specificity and accuracy in digital pathology applications. We also plan to extend testing to other tissue types and applications, broadening the model’s utility and generalisability. Expanding the dataset with more samples and additional staining types would further support these goals.

\section{Conclusions}
\label{section:Conclusion}

This work presents the first known demonstration of 3D virtual staining for X-ray histology, enabled by a modified CycleGAN architecture adapted for paired SR-µCT and histology data of bone-implant specimens. By integrating pixelwise supervision and greyscale loss terms into the traditional CycleGAN framework, our model effectively learns to synthesise colourised virtual histology volumes from greyscale X-ray tomography inputs, bridging the gap between structural and chemical imaging modalities.

Our results show that the modified CycleGAN significantly outperforms both standard CycleGAN and Pix2Pix baselines in terms of structural similarity, perceptual fidelity, and peak signal-to-noise ratio, as confirmed through both qualitative visual comparisons and quantitative metrics (SSIM, PSNR, LPIPS). Crucially, the model not only captures broad tissue structures but also fine histological features, such as bone lacunae and degradation layers, suggesting the potential to support downstream biomedical interpretation.

Applied to the context of biodegradable magnesium-based implants, our method demonstrates the potential of deep-learning-enabled 3D virtual staining to enhance the interpretability of label-free X-ray histology, facilitating scalable, non-destructive, and chemically informative analysis across volumetric datasets. The successful prediction of biologically relevant features such as newly formed bone regions adjacent to implant degradation layers underscores the practical utility of this approach in biomedical research and preclinical evaluation.

While the current model is trained on downscaled whole-slide images and relies on the availability of paired data, its strong performance sets a foundation for future developments. These may include hybrid architectures incorporating unpaired data, improved colour normalisation, upscaled resolutions approaching true histological fidelity, and broader testing across tissue types and staining protocols. Taken together, our findings open the door to integrating virtual staining directly into X-ray imaging pipelines, offering a powerful new tool for digital pathology and 3D biomedical imaging.

\section*{Acknowledgments}

This computational project was carried out with the support of the Joint Laboratory Model and Data-driven Materials Characterisation (JL MDMC), a cross-centre platform of the Helmholtz Association. The work was also supported through use of the Maxwell computational resources operated at Deutsches Elektronen-Synchrotron (DESY), Hamburg, Germany. We thank our summer student  Moral Bootbool for her contribution in the pilot testing of our model. The experimental work is acknowledged extensively within \cite{kruger_high-resolution_2022}. In particular, we are grateful to Silvia Galli for the acquisition of histological images. Parts of this research were supported by the BMBF project ‘Multi-task Deep Learning for Large-scale Multimodal Biomedical Image Analysis (MDLMA)’ (project number 031L0202A), the Hereon project ‘Holistic Data Analysis (HoliDAy)’ of the Innovation-, Information-\& Biologisation-Fonds (I2B), and the ErUM-Data Verbundprojekt ‘KI4D4E: Ein KI-basiertes Framework für die Visualisierung und Auswertung der massiven Datenmengen der 4D-Tomographie für Endanwender von Beamlines’ (project number 05D23CG1) which is funded by the Bundesministeriums für Bildung und Forschung (BMBF).

\bibliography{MyLibrary}

\renewcommand{\appendixname}{Supplementary Information}

\appendix
\renewcommand{\thesection}{S\arabic{section}}    %%%% but here

\section*{Supplementary Information}
\renewcommand{\thefigure}{S\arabic{figure}}
\setcounter{figure}{0}
\renewcommand{\thetable}{S\arabic{table}}
\setcounter{table}{0}

\section{Toluidine blue histology data results}
\label{sec:si_Tblue}

\begin{figure}
\centering
\begin{subfigure}{0.49\textwidth}
    \includegraphics[width=\textwidth]{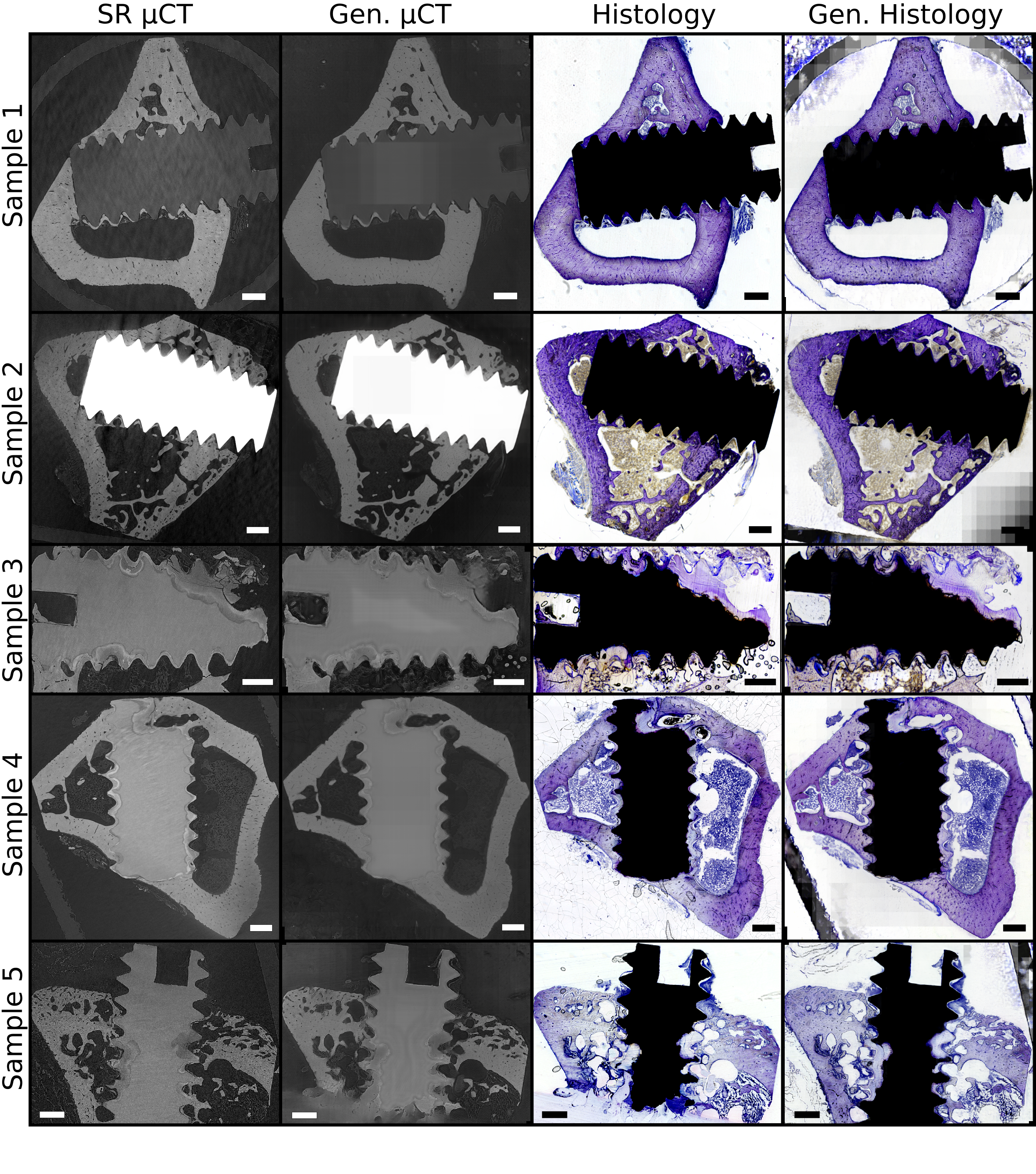}
    \caption{Training results (5 of 40 WSI samples) }
    \label{fig:sup_train_results}
\end{subfigure}
\hfill
\begin{subfigure}{0.49\textwidth}
    \includegraphics[width=\textwidth]{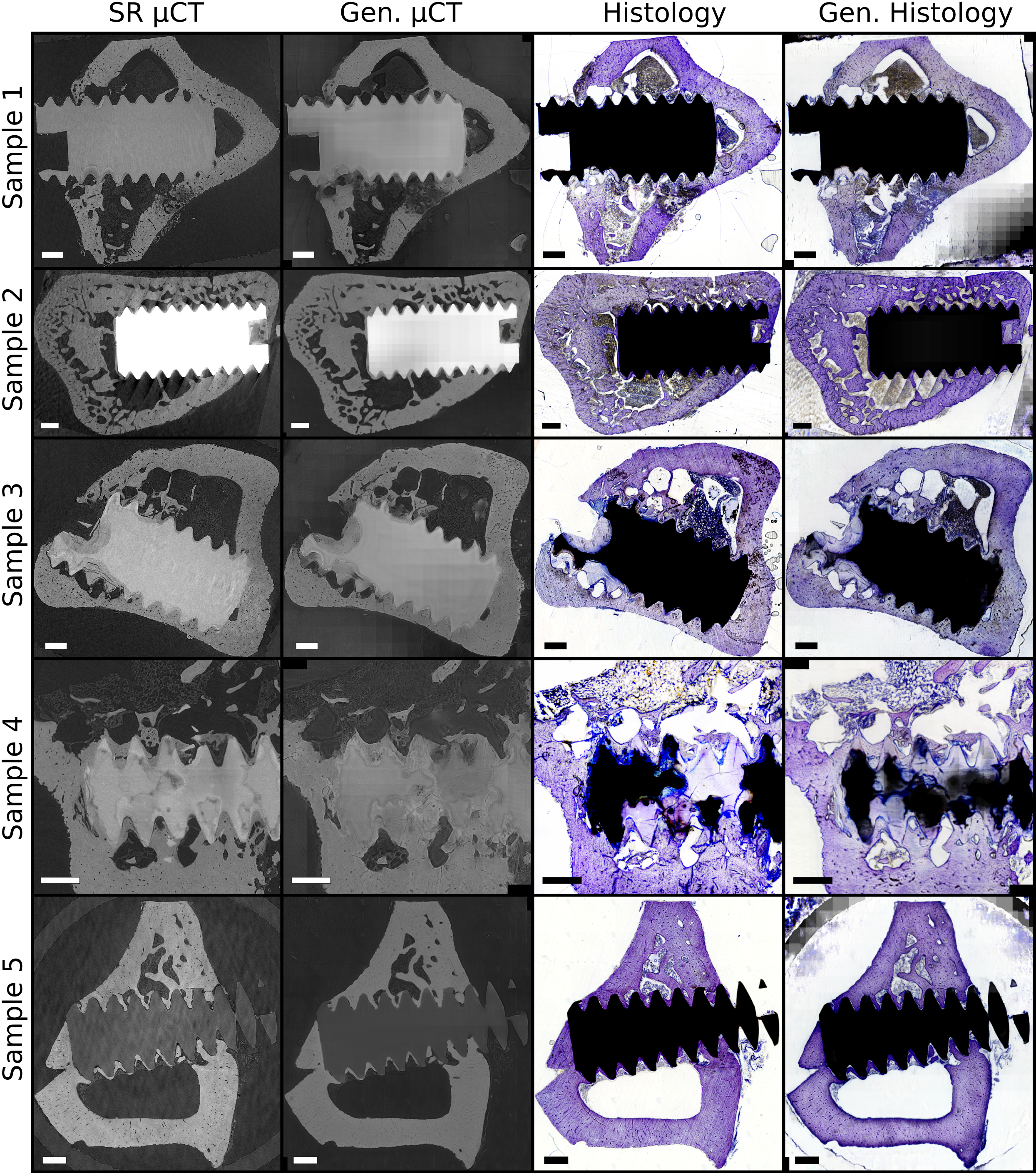}
    \caption{Validation results (5 of 10 WSI samples)}
   \label{fig:sup_val_results}
\end{subfigure}     
\hfill
\hfill
\hfill
\begin{subfigure}{0.49\textwidth}
    \includegraphics[width=\textwidth]{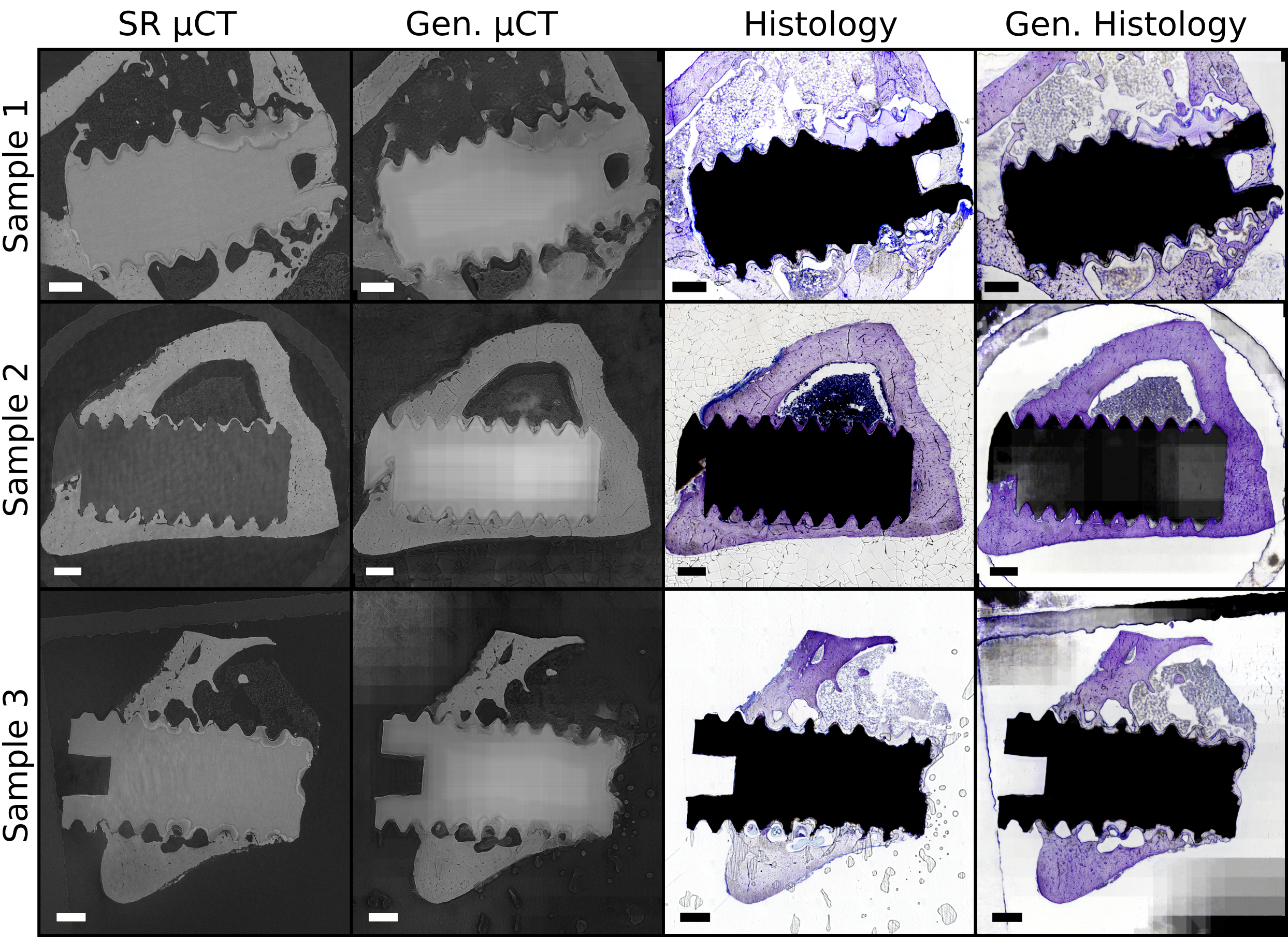}
    \caption{Test results (3 of 3 WSI samples)}
   \label{fig:sup_test_results}
\end{subfigure}    
\caption{The modified CycleGAN model results (with WSI output). In training: Sample 1 = PEEK, Sample 2 = Ti, Sample 3 = Mg, Sample 4 = Mg, Sample 5 = Mg. In validation: Sample 1 = Mg, Sample 2 = Ti, Sample 3 = Mg, Sample 4 = Mg, Sample 5 = PEEK. In testing: Sample 1 = Mg, Sample 2 = PEEK, Sample 3 = Mg. Scalebars represent 500 µm.}
\label{fig:sup_train-test-results}
\end{figure}

A range of examples from each of the training, validation and testing inference results from the modified CycleGAN model are shown in Figures \ref{fig:sup_train_results}, \ref{fig:sup_val_results} and \ref{fig:sup_test_results}. The alloy material of each screw implant is given within the caption. The output WSIs generated through overlapping patch-based inference are displayed without sample masking. Consequently, we can observe various background effects such as tiling artefacts, and predictions based on the walls of the sample holder (seen as straight or circular depending on the cylinder-to-slice geometry).  Sample 1 of each subset were previously shown in Figure \ref{fig:results_validation_testing_profiles}.

\section{H\&E histology data results}
\label{sec:si_HE}

In this section we describe the results of application of our paired models to a smaller, secondary dataset stained with H\&E. 

\subsection{H\&E Datasets}
\label{subsection:si_HEdatasets}

For the secondary dataset stained with H\&E, a total of 11 co-registered µCT and histology partial WSI pairs were able to be collated for this project. These comprised 3 samples containing Mg (-based implants), 6 containing Ti and 2 containing PEEK. We used the same hyperparameters for the modified CycleGAN and Pix2Pix as determined through validation performed on the Toluidine Blue dataset, but the models were re-initialised from scratch without pre-training before introducing the H\&E dataset. For this qualitative demonstration, training was performed on 8 WSI pairs (2 Mg, 5 Ti and 1 PEEK) with the remaining 3 pairs for validation/testing (1 each of Mg, Ti and PEEK). 

Due to the laser cutting process for this dataset (as opposed to the physical cutting and grinding of the Toluidine Blue (see Section \ref{subsection:SamplePreparation}) the residual screw alloy components are missing from the histology slices. Consequently the screw appears white instead of black (with the one exception of Sample 2 in Figure \ref{fig:sup_red_testing}, which retains a half section of residual Mg-based alloy in black). There are a few missing sections of tissue too, resulting also in corresponding white regions. The fields of view (FOV) of these partial WSIs are limited and non-rectangular; regions outside of the existing stitched data are depicted as black.  Sample correspondence maps were created to reflect the stitched FOVs. 

\subsection{Description of H\&E  results}
\label{subsection:si_HEdescription}

\begin{figure}
\centering
\begin{subfigure}{0.49\textwidth}
    \includegraphics[width=\textwidth]{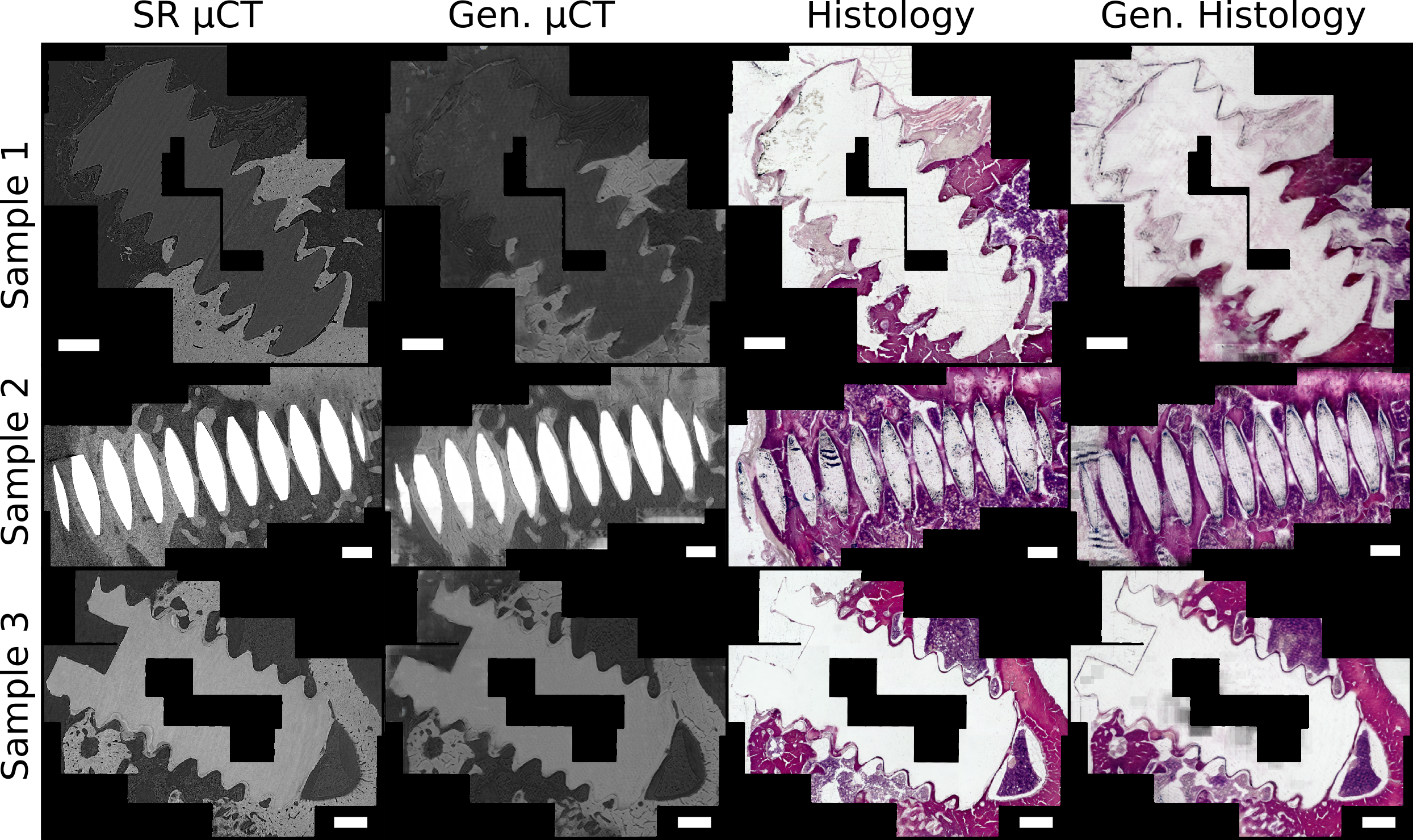}
    \caption{training results (3 of 8 WSI samples)}
    \label{fig:sup_red_training_masked}
\end{subfigure}
\hfill
\begin{subfigure}{0.49\textwidth}
    \includegraphics[width=\textwidth]{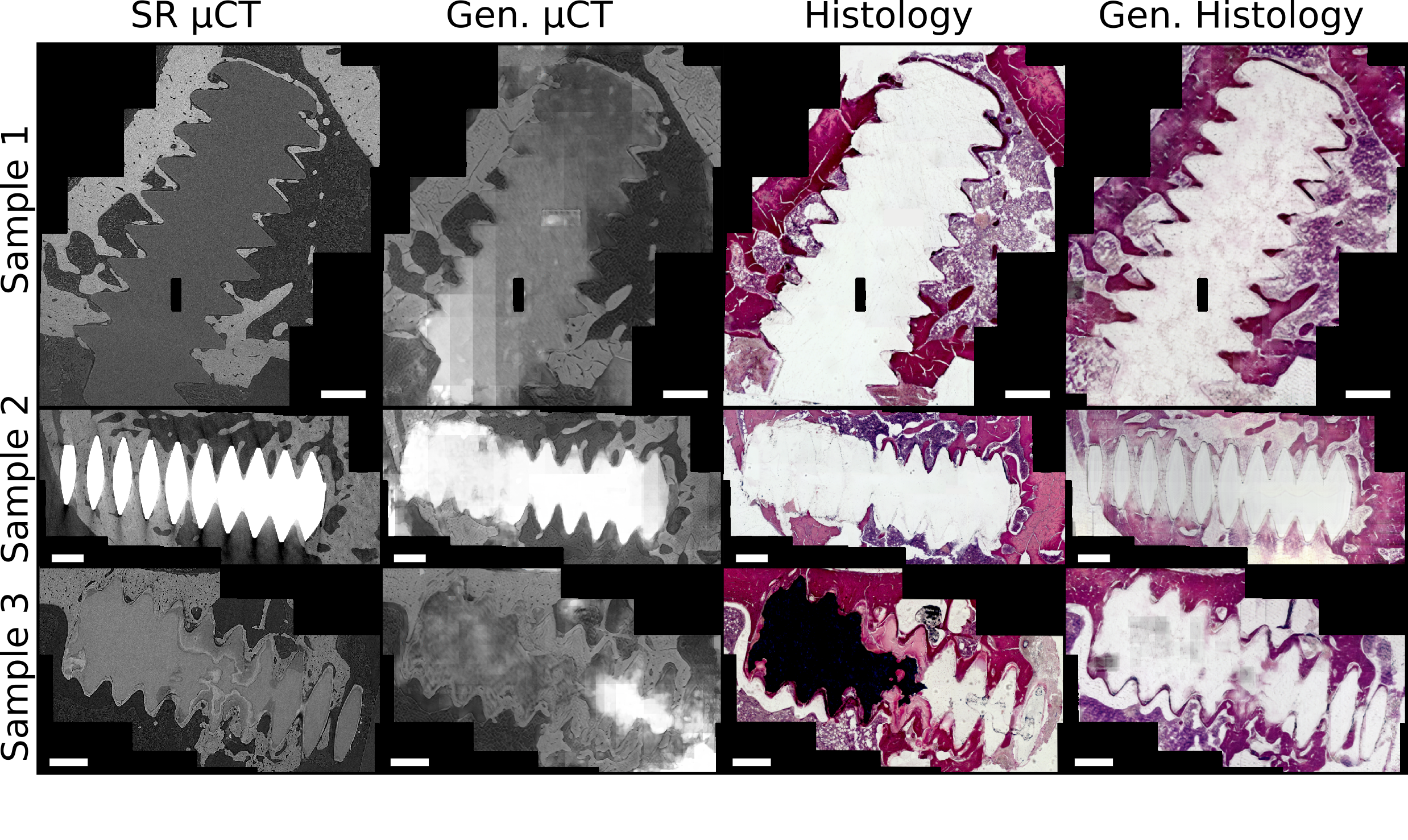}
    \caption{validation/testing results (3 of 3 WSI samples)}
    \label{fig:sup_red_testing_masked}
\end{subfigure}        
\caption{Masked views of the inference results of the modified CycleGAN model applied to the secondary H\&E stained dataset (masked view). In training: Sample 1 = PEEK, Sample 2 = Ti, Sample 3 = Mg. In validation/testing: Sample 1 = PEEK, Sample 2 = Ti, Sample 3 = Mg. Scalebars represent 500 µm.}
\label{fig:sup_red_results_masked}
\end{figure}

\begin{figure}
\centering
\begin{subfigure}{0.49\textwidth}
    \includegraphics[width=\textwidth]{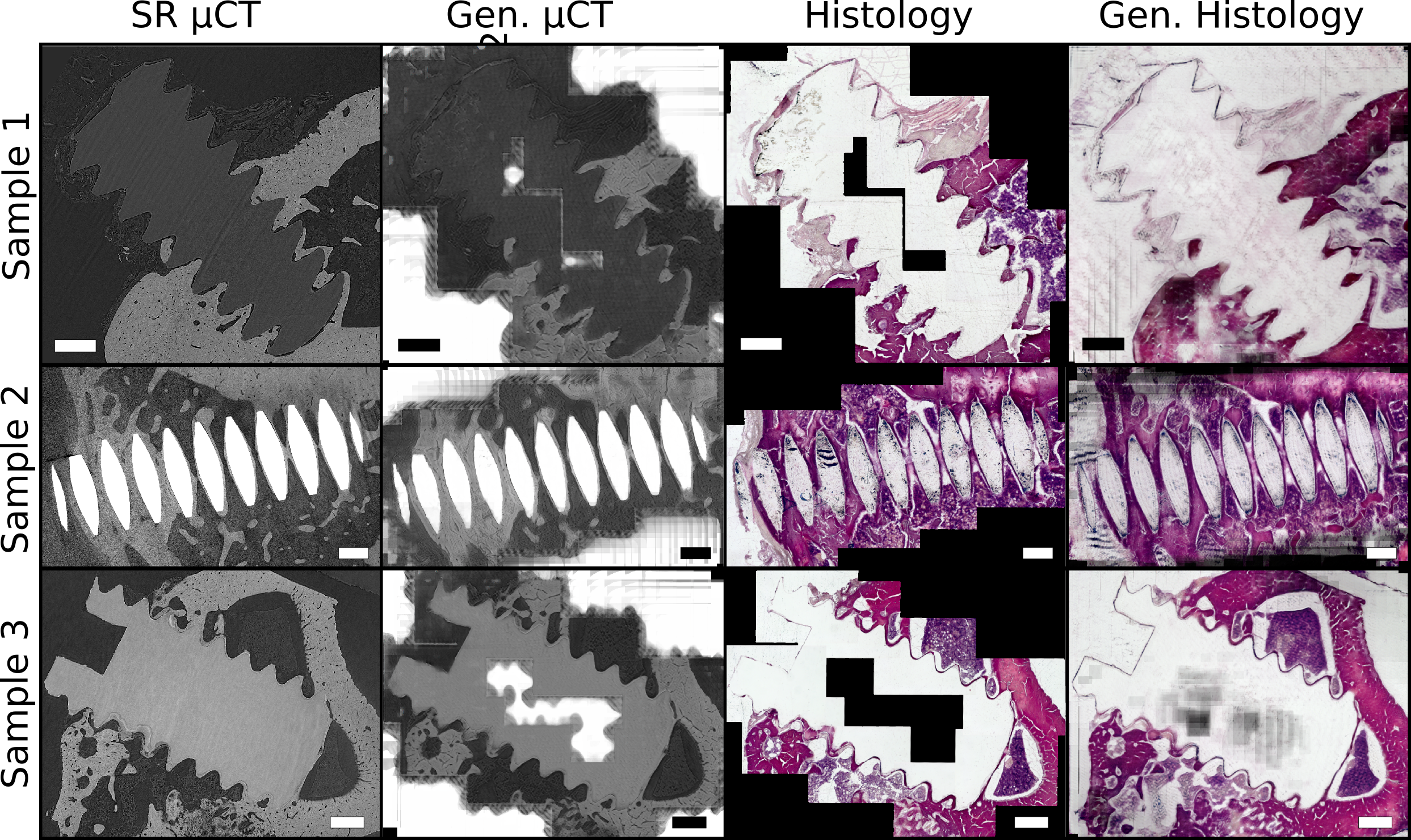}
    \caption{training results (3 of 8 WSI samples)}
    \label{fig:sup_red_training}
\end{subfigure}
\hfill
\begin{subfigure}{0.49\textwidth}
    \includegraphics[width=\textwidth]{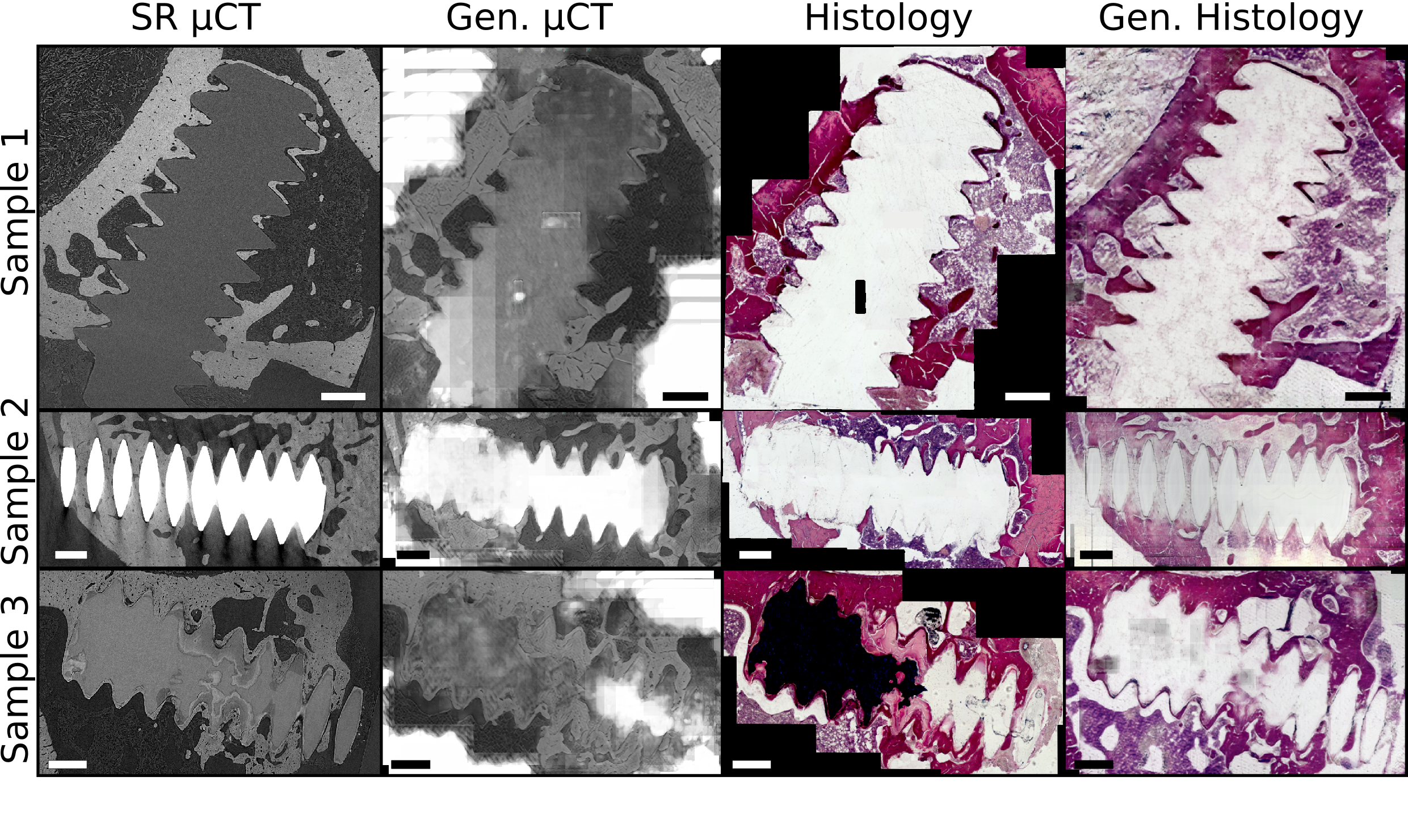}
    \caption{validation/testing results (3 of 3 WSI samples)}
    \label{fig:sup_red_testing}
\end{subfigure}        
\caption{Unmasked views of the inference results of the modified CycleGAN model applied to the secondary H\&E stained dataset. In training: Sample 1 = PEEK, Sample 2 = Ti, Sample 3 = Mg. In validation/testing: Sample 1 = PEEK, Sample 2 = Ti, Sample 3 = Mg. Scalebars represent 500 µm.}
\label{fig:sup_red_results}
\end{figure}

Three examples from each of the training, and validation/testing inference results from the modified CycleGAN model are shown in Figures \ref{fig:sup_red_results_masked} (masked view) and \ref{fig:sup_red_results} (unmasked view). A direct comparison is more easily made with the masked view. In the unmasked view, the incompleteness of the tiled field of view causes uncertainty outside of these stitched regions (since the sample correspondence mask is also applied to the loss terms during training). This yields false areas of white in the generated µCT images and black in the histology. On the other hand, the generated histology images are reasonably able to be predicted across the full rectangular FOV provided by the input µCT slice. 

In Figure \ref{fig:sup_red_results_masked}, we see a promising level of agreement between real and generated histology images in terms of structure and colour. 
Although the dataset is small, these results demonstrate that our modified CycleGAN model may be reasonably applied to larger H\&E datasets as well as Toluidine Blue, suggesting its utility as a general stain-appropriate model. 
However, it is not really possible to judge the interpretability of the trained model results in terms of previously mentioned features such as degradation layer and new bone formation for a number of reasons. Firstly, the histology quality is not as good, particularly surrounding the region which contained the screw implant in several samples (see for example the missing tissue sections in Sample 2 of Figure \ref{fig:sup_red_testing_masked}.) Secondly, the degradable samples were here under-represented, with only 3 total Mg samples present in which to observe the degradation layers. In those samples, the H\&E stain differentiates less strongly between the degradation layer, new bone and even some parts of soft tissue, all of which are observed to present in various shades of pale pink, as compared to the deep pink-red of the dense bone material. 

\subsection{Qualitative comparison of our model vs Pix2Pix}
\label{subsection:si_HEmodels}

\begin{figure}
\centering
\includegraphics[width=0.7\textwidth]{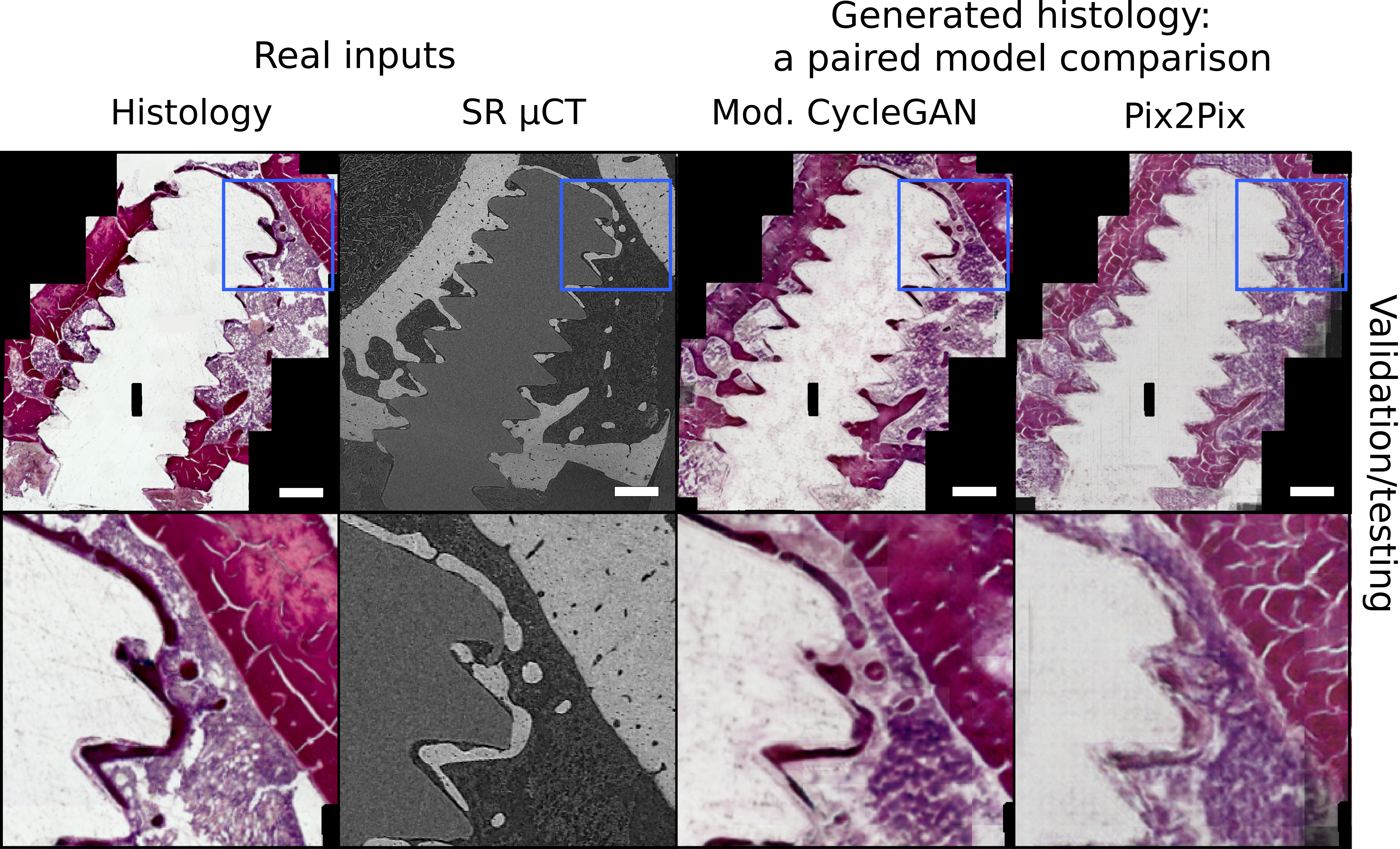}
\caption{Comparison of inference results from the modified CycleGAN and Pix2Pix paired model with the H\&E stained data validation/testing set. Including masked WSI and 256 $\times$ 256 pixel ROI examples. The scalebar in each WSI represents 500 µm.}
\label{fig:sup_model_comparison_HE}
\end{figure}

In Figure \ref{fig:sup_model_comparison_HE} we qualitatively compare the H\&E stain-based inference results of our paired CycleGAN model with those of Pix2Pix, with an example validation/testing WSI plus $256 \times 256$ pixel ROI. The generated histology images are shown as masked with the sample correspondence map.

It is clear that our model-generated histology image is a more accurate prediction of the real histology image. In both models, the soft tissue (coloured purple) is not very sharply reproduced, but they are better defined by the CycleGAN model. Pix2Pix also fails to give a strong delineation between the region belonging to the implant (here PEEK-based, which has osseointegrated) and the surrounding bone. Interestingly, the Pix2Pix generated regions of dense bone feature a high degree of style-based realism in the way it has reproduced the white cracks characteristic of the real histology dense bone. However, these cracks were formed in the bone sections after the whole volume tomography acquisition process and do not correspond to the real µCT data. These may be classed as hallucinations by the GAN, produced in an effort to satisfy the discriminator.  

\section{Relative resolution of the generated histology images}
\label{subsection:sup_resolution}

\begin{figure}
\centering
\begin{subfigure}{0.45\textwidth}
    \includegraphics[width=\textwidth]{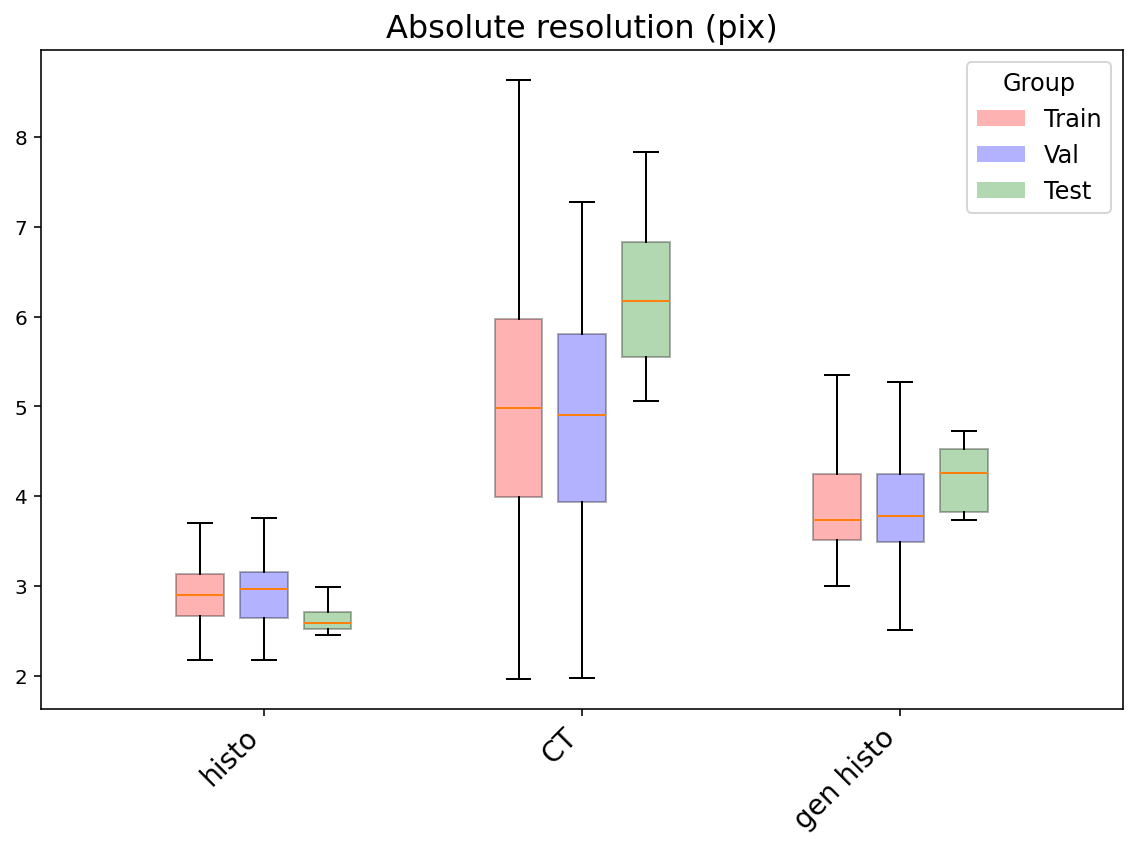}
    \caption{}
    \label{fig:abs-frc}
\end{subfigure}
\hfill
\begin{subfigure}{0.45\textwidth}
    \includegraphics[width=\textwidth]{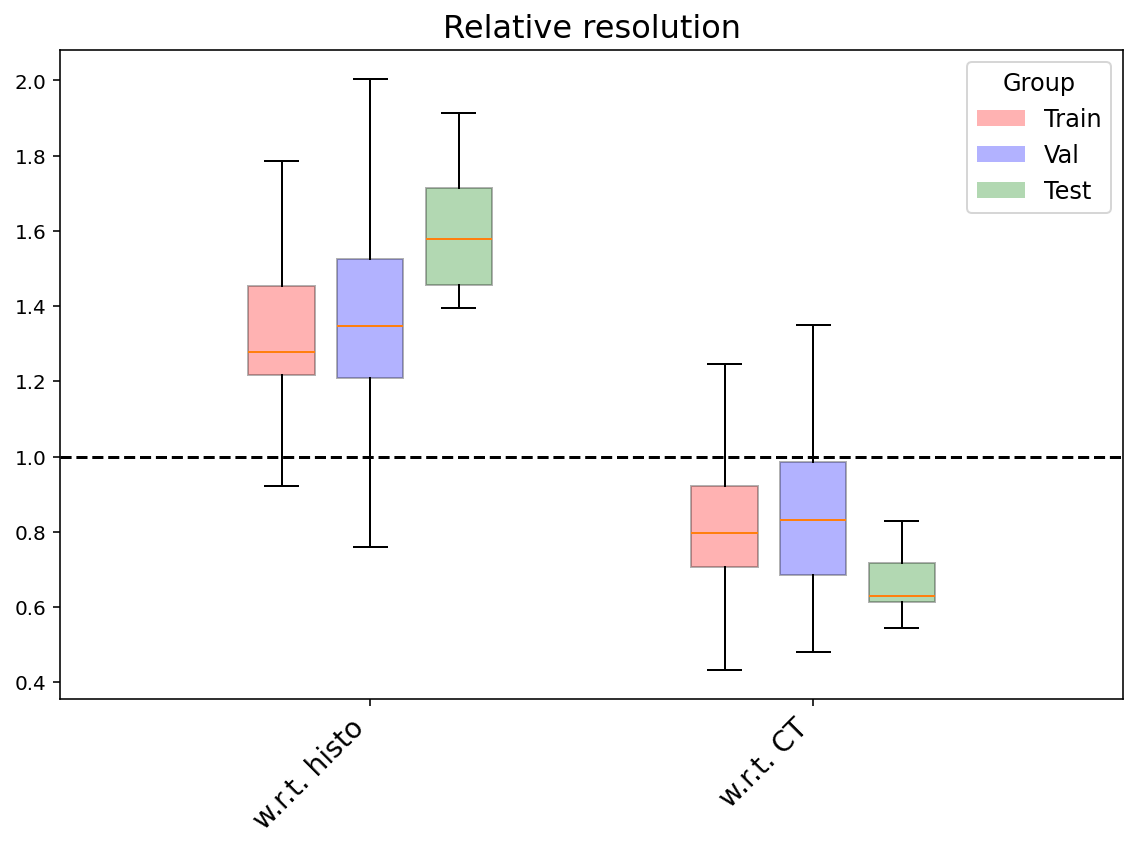}
    \caption{}
    \label{fig:rel-frc}
\end{subfigure}        
\caption{Box plots of the measured resolution of the generated histology compared to the input real histology and CT data, shown in both absolute values (in units of pixels) in (a), and in relative terms in (b).}
\label{fig:sup_res_frc}
\end{figure}

We performed preliminary image spatial resolution measurements on the input/output Toluidine Blue-stained data, in a study of our modified CycleGAN model. Measurements were based on the Fourier ring correlation method \citep{nieuwenhuizen_measuring_2013, rieger_single_2024}, and sampling was performed with 256 $\times$ 256 pixel patches across the WSI images, excluding the masked areas. Absolute values in units of pixels were calculated for both the generated histology images as well as the original input CT and histology images, from which we also derived a spatial resolution value relative to both inputs. The relative fraction was calculated per image pair due to high levels of image inhomogeneity. The results of both are shown with two plots in Figure \ref{fig:sup_res_frc}, for each of the train, validation and test groups. In absolute resolution terms, the input CT was generally considerably lower in resolution than the corresponding input histology (with a mean of around 5-6 pixels, compared to less than 3 pixels for the histology), whilst the generated histology resolution was somewhere in-between (around 4 pixels). In relative terms, for the training and validation sets the generated histology resolution was found to be on average 1.35$\times$ the resolution of the histology and 0.8 $\times$ the resolution of the CT. In the three test WSI pairs, the input CT images were notably worse in resolution than the 50 train+validation pairs, and this effect became more pronounced (with a value of less than 0.7 $\times$ the resolution of the CT).  Note however that it is difficult to accurately measure the resolution due to various high-frequency artefacts present in the images which may influence the result. It is also not straightforward to directly compare the signal-to-noise distributions of the histology and CT images since they are so different. A more comprehensive study may be performed at a later stage, perhaps in combination with a higher-resolution dataset (involving less-downsampling of the original WSI histology images). 

\end{document}